\documentclass[reprint, superscriptaddress, prl]{revtex4-2}

\usepackage{siunitx}
\usepackage[english]{babel}
\usepackage[utf8]{inputenc}

\usepackage{float}
\usepackage{graphicx}   
\graphicspath{{/figures}}

\usepackage{amsmath}
\usepackage{amsfonts}
\usepackage{relsize}
\usepackage{empheq}
\usepackage{braket}
\usepackage{bbm}
\usepackage{bm}
\usepackage{cancel}

\usepackage[section]{placeins}
\usepackage{xcolor}
\usepackage{layouts}

\usepackage[colorlinks, hypertexnames=false]{hyperref}
\definecolor{navygray}{RGB}{110,140,170}
\hypersetup{
citecolor={navygray},
linkcolor={navygray},
urlcolor={navygray},
}
\usepackage[all]{hypcap}

\newcommand{\eref}[1]{\hyperref[#1]{{Eq.~\ref{#1}}}}  
\newcommand{\eqsref}[1]{\hyperref[#1]{{Eqs.~\ref{#1}}}}  

\newcommand{\fref}[1]{\hyperref[#1]{{Fig.~\ref{#1}}}}
\newcommand{\frefadd}[2]{\hyperref[#1]{{Fig.~\ref*{#1}#2}}}

\newcommand{\sref}[1]{\hyperref[#1]{{Sec.~\ref{#1}}}}

\begin{document}

\title{The Bragg demagnifier: X-ray imaging with kilometer propagation distance \newline within a meter}

\author{Rebecca Spiecker}
\affiliation{LAS,~Karlsruhe~Institute~of~Technology,~76131~Karlsruhe,~Germany}
\affiliation{IPS,~Karlsruhe~Institute~of~Technology,~76344~Eggenstein-Leopoldshafen,~Germany}

\author{Martin Spiecker}
\affiliation{IQMT,~Karlsruhe~Institute~of~Technology,~76344~Eggenstein-Leopoldshafen,~Germany}
\affiliation{PI,~Karlsruhe~Institute~of~Technology,~76131~Karlsruhe,~Germany}

\author{Adyasha Biswal}
\affiliation{LAS,~Karlsruhe~Institute~of~Technology,~76131~Karlsruhe,~Germany}
\affiliation{IPS,~Karlsruhe~Institute~of~Technology,~76344~Eggenstein-Leopoldshafen,~Germany}
\affiliation{COS,~Heidelberg~University,~69120~Heidelberg,~Germany}

\author{Mykola Shcherbinin}
\affiliation{LAS,~Karlsruhe~Institute~of~Technology,~76131~Karlsruhe,~Germany}

\author{Tilo Baumbach}
\affiliation{LAS,~Karlsruhe~Institute~of~Technology,~76131~Karlsruhe,~Germany}
\affiliation{IPS,~Karlsruhe~Institute~of~Technology,~76344~Eggenstein-Leopoldshafen,~Germany}

\date{\today}

\begin{abstract}

We introduce a new X-ray imaging technique to facilitate propagation-based phase contrast of large, centimeter-sized samples. The diffracted X-ray wavefield behind the sample is demagnified by asymmetric Bragg crystal optics, thereby virtually increasing the propagation distance and thus enhancing the image contrast. We demonstrate the significant increase in image contrast compared to conventional phase contrast imaging at the same short physical propagation distance. Additionally, the Bragg demagnifier enables the reduction of image blur caused by the finite X-ray source size. In combination with a subsequent Bragg magnifier, the method will allow for an even higher dose efficiency, rendering this technique a potential candidate for, e.g., low-dose (bio)medical diagnostics.

\end{abstract}

\maketitle

Conventional X-ray imaging relies on the attenuation of X-rays in a medium and is well suited to highly-absorbing materials, but yields only vanishingly small contrast for soft materials and tissues~\cite{Salditt2017Oct}. Phase contrast imaging can overcome these limitations by uncovering the phase shift that the sample imprints on the X-ray wavefield~\cite{Salditt2017Oct,endrizzi2018x}. In the last decades, multiple phase contrast imaging techniques have been developed~\cite{Chapman2010Dec,Pfeiffer2013Sep,Lider2013Nov,endrizzi2018x,Tao2021Mar}, finding application in various fields such as materials science~\cite{mayo2012line}, natural and cultural heritage~\cite{cotte2019cultural}, biology~\cite{Betz2007Jul}, industrial and medical applications~\cite{kastner2017new,lewis2004medical,Bravin2012Dec, Taba2018May}.
\vspace*{1.82ex}
\newline \indent For phase contrast imaging of large, centimeter-sized samples at moderate resolution, it is essential to attain contrast at low spatial frequencies. This holds in particular for flux-limited applications such as low-dose or fast imaging, where a compromise has to be taken between deposited dose or exposure time on the one hand, and spatial resolution on the other hand.
There exist two main approaches for phase contrast imaging of large samples at moderate resolution above ${\sim}\,\SI{10}{\micro m}$: differential phase contrast, e.g. analyzer-~\cite{davis1995phase, chapman1997diffraction}, grating-~\cite{David2002Oct,momose2003demonstration, Pfeiffer2013Sep} or speckle-based~\cite{Morgan2012Mar,Zdora2018Apr}, and propagation-based phase contrast imaging (PB-PCI)~\cite{snigirev1995possibilities, Wilkins1996Nov, cloetens1999holotomography,zamir2016x,wagner2018towards,donnelley2019live,walsh2021imaging}. 
PB-PCI utilizes the coherent self-interference of the transmitted wavefield behind the sample, which gradually evolves into intensity contrast with increasing propagation distance.
However, to visualize low spatial frequencies of ${\sim}\,2\pi/\SI{10}{\micro m}$ to ${\sim}\,2 \pi / \SI{1}{mm}$, long propagation distances of ideally tens to hundreds of meters are necessary~\cite{zamir2016x, wagner2018towards, donnelley2019live, walsh2021imaging, tavakoli2021tracking, arhatari2021x, cotte2022new, haggmark2023phase}. Recently, a new beamline has been built at the European Synchrotron Radiation Facility (ESRF, Grenoble, France) with a remarkably long experimental hutch of \SI{45}{m}, which facilitates PB-PCI at propagation distances of up to \SI{36}{m}, tailored to the X-ray source size~\cite{Cianciosi2021Oct, cotte2022new, Lang2023Mar}.
\vspace*{1.82ex}
\newline \indent Here, we describe a new imaging technique that enables PB-PCI at moderate resolution with a physical distance between sample and detector in the sub-meter range, yet allowing for long \textit{effective} propagation distances of hundreds to thousands of meters. For this purpose, the spatial frequency distribution of the diffracted X-ray wavefield behind the sample is magnified by asymmetric Bragg diffraction, resulting in a strongly increased effective propagation distance and thus increased image contrast. In the following, we outline the basic working principle, describe the realization by Bragg crystal optics and show a first proof-of-concept experiment.\\

\begin{figure*}[t!]
\begin{center}
\includegraphics[scale=1.0]{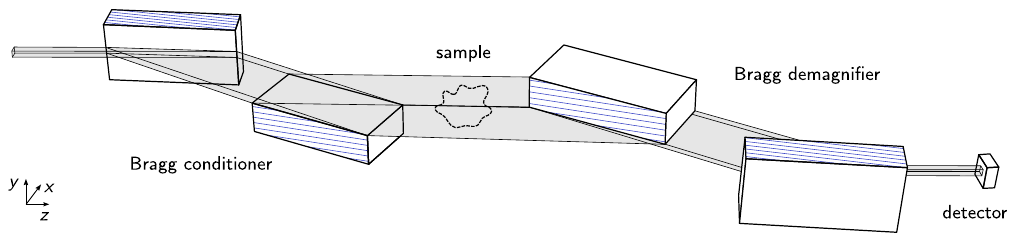}
\caption{Schematic of the crystal arrangement. The first two crystals (Bragg conditioner) magnify the incoming X-ray beam cross section in horizontal and vertical direction. The diffracted wavefield behind the sample is then demagnified in vertical and horizontal direction (Bragg demagnifier), ideally using the same pair of crystals as for the Bragg conditioner. After free space propagation of the demagnified wavefield, we record the image wavefield with a high-resolution scintillator-based detector. 
Alternatively, the X-ray wavefield may again be magnified to be recorded efficiently, e.g., with a single photon counting detector~\cite{Spiecker}. The blue lines illustrate the crystal lattice planes.
} \label{fig_1}
\end{center}
\end{figure*}

The working principle is best understood in Fresnel diffraction and assuming monochromatic plane wave illumination of the sample. The complex-valued wavefield $f(\mathbf{r})$ in the sample exit plane at $z=0$ can be represented by its Fourier transform $\tilde{f}(\mathbf{k})$, where $\mathbf{r} = (x,y)^T$ and $\mathbf{k}=(k_x,k_y)^T$ denote the spatial coordinate and angular spatial frequency, respectively.
After free space propagation along a distance~$z$, the wavefield $g(\mathbf{r})$ in the image plane is given by 
\begin{align}\label{eq_1}
g(\mathbf{r}) = \frac{1}{2 \pi}\iint \tilde{f}(\mathbf{k}) e^{-i\frac{\mathbf{k}^2}{2k_0} z} e^{i \mathbf{k}\mathbf{r}}\text d k_x \text d k_y,
\end{align}
where $k_0$ is the angular wave number in vacuum and $e^{-i\frac{\mathbf{k}^2}{2k_0} z}$ is denoted as the Fresnel propagator~\cite{Cowley1978Jan}. The argument $\frac{\mathbf{k}^2}{2k_0} z$ shows that low spatial frequencies in the wavefield need a large propagation distance to accumulate a certain phase shift, which can be converted into measurable intensity contrast. For imaging large samples at moderate resolution, it would thus be desirable to magnify the spatial frequencies of the sample exit wavefield in order to avoid tens to hundreds of meters physical propagation distance. By magnifying the spatial frequencies by a factor $M$, the resulting wavefield $\tilde{f}'$ is given by $\tilde{f}'(\mathbf{k}) = \tilde{f}(\mathbf{k} / M)$.
Substitution of $\mathbf{k} / M \rightarrow \mathbf{k}$ yields the wavefield
\begin{align}\label{eq_1b}
g'(\mathbf{r}) = \frac{M^2}{2 \pi}\iint \tilde{f}(\mathbf{k}) e^{-i\frac{{\mathbf{k}}^2}{2k_0} M^2 z} e^{i \mathbf{k} M\mathbf{r}}\text d k_x \text d k_y
\end{align}
in the image plane after propagation of the modified wavefield along the physical distance $z$.
The expansion of the frequency spectrum thus leads to an enhanced effective propagation distance $z_\text{eff} = M^2 z$ that is quadratic in $M$. The image acquired at the physical distance $z$ will therefore exhibit the same contrast as the unaltered wavefield after the propagation distance $M^2z$. By the term $M \mathbf{r}$ in \eref{eq_1b} we recognize that the magnification of the spatial frequencies comes along with a demagnification in real space. 
The demagnified image can either be detected by a high-resolution detector, which, however, has a limited detection efficiency, or it can be subsequently re-magnified for being detected with an efficient large-area detector~\cite{Spiecker}, see also Appendix.
\\

Demagnifying the X-ray wavefield with minimal distortion can be achieved by using a so-called Bragg magnifier with reversed optical path (see \fref{fig_1}, right). A Bragg magnifier magnifies an X-ray wavefield by asymmetric Bragg diffraction~\cite{boettinger1979x, kobayashi2001x, stampanoni2002bragg, modregger2006magnified}. In this case, the asymmetry angle $\alpha$ between the crystal surface and the reflecting lattice planes is defined as positive. In contrast, the wavefield is compressed when the asymmetry angle is negative. So far, demagnification by asymmetric Bragg diffraction has been used for focussing and collimating the incident X-ray illumination~\cite{watanabe1999rotated,tsusaka2000formation}, to achieve a high angular sensitivity in analyzer-based imaging~\cite{davis1995phase} or to fit a large X-ray image onto a smaller detector~\cite{hirano2013applications}. Here, we use demagnification to specifically increase image contrast in PB-PCI.
In our case, the Bragg demagnifier consists of two crystals for demagnification in $x$- and $y$-direction, respectively. 
The dependency between the spatial frequencies $k_\text{in}$ and $k_\text{out}$ of the wavefield in front of and behind the crystals is in general a slightly nonlinear function (see Appendix), causing the image formation to be shift-variant. In good approximation and to keep the discussion simple, we have $k_\text{out}(k_\text{in}) \approx M k_\text{in}$, restoring the shift-invariance. 
As for the Bragg magnifier~\cite{stampanoni2002bragg}, the demagnification factor $M$ is given by
\begin{align*}
    M =
\frac{\sin(\theta_\text{B} + \Delta\theta_\text{oc} - \alpha)}{
\sin(\theta_\text{B} + \Delta\theta_\text{hc} + \alpha)}.
\end{align*}
Here, $\alpha < 0$ is the asymmetry angle, $\theta_\text{B}$ the Bragg angle and $\Delta\theta_\text{oc}$ and $\Delta\theta_\text{hc}$ are refraction correcting terms from dynamical diffraction theory for the incident and outgoing wavefield, respectively~\cite{authier2004dynamical}. 
\\

Since the two crystals cannot be physically positioned at the same location but have a certain distance to each other, the horizontal and vertical compression of the wavefield occurs at different distances $z_{1,x}$ and $z_{1,y}$ along the optical axis between the sample and crystal centers. Likewise, the distances $z_{2,x}$ and $z_{2,y}$ between the crystal centers and the detector are different as well. The demagnified wavefield in the detector plane is given by
\begin{align*}
    g'(\mathbf{r}) &= \frac{M_xM_y}{2\pi}\iint \tilde{f}(\mathbf{k})H(\mathbf{k})\,e^{ik_xM_x x}e^{ik_yM_y y}\text d k_x \text d k_y
    \intertext{with the propagator}
    H(\mathbf{k}) &= R(\mathbf{k}) e^{-i\frac{k_x^2}{2k_0}(z_{1,x}+M_x^2z_{2,x})} e^{-i\frac{k_y^2}{2k_0}(z_{1,y} + M_y^2z_{2,y})}.
\end{align*}
In $R(\mathbf{k})$ we have combined the complex crystal reflection of the diffracted wavefield amplitudes of both crystals~\cite{authier2004dynamical}. $M_x$ and $M_y$ are the demagnification factors in horizontal and vertical direction, respectively. The different distances and potentially different magnifications lead to different effective propagation distances 
\begin{align*}
z_{\text{eff},i}=z_{1,i}+M_i^2z_{2,i}
\end{align*}
in horizontal and vertical direction. For large $M$, the propagation along $z_{1,i}$ in front of the crystals, i.e. before demagnification, can be neglected. Lastly, the highest resolvable spatial frequency $k_\text{max}$ is restricted by the angular acceptance of the crystals, i.e. the input Darwin width $\delta_\text{oc}$~\cite{authier2004dynamical}. In accordance to Abbe's criterion, it holds $k_\text{max} = k_0 \delta_\text{oc}$~\cite{abbe1873beitrage}.
\vspace{1.083ex}
\newline \indent A prerequisite for imaging with a Bragg demagnifier is a monochromatic X-ray beam with sufficient transversal coherence and a large beam cross section to illuminate the sample. 
In combination with a double-crystal monochromator, synchrotron beamlines of third and fourth generation provide low divergence and high monochromaticity. However, they usually offer only a limited, millimeter-sized beam cross section and finite transversal coherence at the sample position~\cite{Martinson2015May,markus2018optimizing,reinhard2021beamline}. 
Here, we magnify the beam cross section in front of the sample by a Bragg magnifier, which we denote as Bragg conditioner (see \fref{fig_1} left)~\cite{Koch1983,christensen1994expanded,Kamezawa2023Sep}. In addition, this Bragg conditioner further increases the transversal coherence and reduces the monochromatic beam divergence to well below the incident angular acceptance of the demagnifier.
Besides, such a Bragg conditioner has the ability to counteract energy dispersion effects of the Bragg demagnifier, as we will revisit later. 
\vspace{1.083ex}
\newline \indent 
As a proof-of-concept, we realize a Bragg demagnifier for an energy range of \SIrange[range-phrase=--,range-units=single]{29}{31}{keV}. A schematic of the crystal arrangement is given in \fref{fig_1}. The incident X-ray beam is expanded in two dimensions by a Bragg conditioner, traverses the sample, is demagnified by the Bragg demagnifier and propagates to the detector plane. We detect the final X-ray image with a high-resolution scintillator-based indirect detector (for details see Appendix). The vertical demagnification is performed first since the source size in third-generation synchrotrons is smaller in vertical direction than horizontally, allowing for larger propagation distances before the onset of source blur. The detector is placed at a distance of $z_{2,y}=\SI{0.90}{m}$ from the first crystal and $z_{2,x}=\SI{0.57}{m}$ from the second crystal. All crystals are composed of silicon (Si) and have an asymmetry angle of $\alpha=\pm \SI{5.92}{\degree}$ between the crystal surface normal and the crystallographic [110] direction. We operate the system at \SI{29}{keV}, where the demagnification factor for the Si 220 reflection is $M=25.7$, resulting in effective propagation distances of $z_{\text{eff},x}\approx\SI{380}{m}$ and $z_{\text{eff},y}\approx\SI{600}{m}$. The nominal resolution limit as introduced above is \SI{68}{\micro m}. The effective pixel size in the detector amounts to \SI{37}{\micro m}. For comparison to conventional free space propagation at the same physical distance, we acquire images without the demagnifier and using a large area detector of similar pixel size (\SI{49.5}{\micro m}, see Appendix), placed at \SI{1}{m} from the sample.
\begin{figure}[t!]
\begin{center}
\includegraphics[scale=1.0]{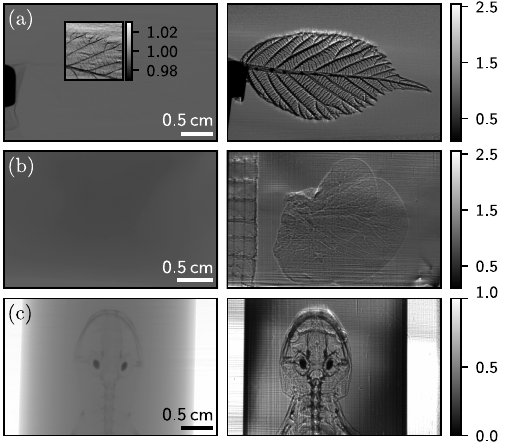}
\caption{Comparison of phase contrast imaging using conventional free space propagation (left) and the Bragg demagnifier (right). In both setups the incident X-ray beam was expanded in front of the sample using a Bragg conditioner~\cite{Davis1995Apr} and the physical propagation distance was ${\sim}\,\!\SI{1}{m}$. For the Bragg demagnifier, this corresponds to an effective propagation distance of $z_{\text{eff},x} \approx \SI{380}{m}$ and $z_{\text{eff},y} \approx \SI{600}{m}$ in $x$- and $y$-direction, respectively, achieved by a demagnification of $M = 25.7$.
(a)~Blackberry leaf in air.  In the conventional free space propagation image only marginal absorption contrast can be seen, as shown in the inset with adjusted gray scale. The Bragg demagnifier provides strong phase contrast.
(b)~Mouse liver lobe in ethanol. Conventional PB-PCI does not reveal any contrast, whereas the Bragg demagnifier brings to light a net of blood vessels. The tape left to the sample becomes visible as well.
(c)~Salamander in ethanol. The Bragg demagnifier yields strong image contrast and allows to visualize soft tissues, while conventional PB-PCI yields only little absorption and no phase contrast. The colorbar indicates the intensity values of the flat- and dark-field corrected images.
} \label{fig_2}
\end{center}
\end{figure}
\vspace{1.083ex}
\newline \indent In \fref{fig_2}, we display example images (after flat- and dark-field correction) of a blackberry leaf, a mouse liver lobe in an ethanol bag and a salamander in a tube filled with ethanol. For conventional PB-PCI, the leaf exhibits only very weak contrast caused by absorption in the sample, as shown in the inset, while the leaf veins become well visible with the demagnifier. The liver is not discernible at all in the conventional image. In contrast, the demagnifier brings to light a net of blood vessels due to the strongly increased propagation distance. A similar increase in contrast is observed for the salamander: in the conventional PB-PCI image, only the bones are weakly visible while the demagnifier reveals the internal structures. The stripes that appear mainly in the images of the liver lobe and the salamander result from crystal polishing artifacts in combination with drifts in the flat-field (empty beam). 
The mechanical stability of our setup made the operation at an even higher magnification of $M = 70$ and thus even larger propagation distances of $z_\text{eff}\approx\SI{4}{km}$ challenging, since the demagnifier crystals have a very small angular acceptance (${\lesssim}\,\SI{1}{\micro rad}$). For routine applications, the stability has to be further improved.
\\

The images acquired with the Bragg demagnifier certainly show patterns that are brighter than the background, indicating constructive interference generated by PB-PCI. However, at the edges of the crystals where the background intensity is reduced, we observe also other bright patterns after flat-field correction, e.g. at the upper edge of the leaf and at the mouth of the salamander. These artifacts arise most likely from the curvature of the incident wavefield, which, at the crystal edges, deviates from the center of the crystal reflectivity curve, thereby generating diffraction-enhanced contrast~\cite{chapman1997diffraction}. In a future setup, the curvature may be reduced by collimating the X-ray beam in front of the Bragg conditioner by compound refractive lenses~\cite{Shastri2004Mar}. Further, the presented images partially contain extinction contrast as known from analyzer-based imaging~\cite{bravin2003exploiting}, which occurs when X-rays are scattered to angles higher than the crystal's angular acceptance and are thus not transferred into the image. We attribute for example the dark spots in the salamander, presumably in the otic region, not only to absorption but also to extinction contrast, since their intensity is reduced with the demagnifier. Extinction contrast is also observed in the bones. 
Note that even though the angular acceptance of the demagnifier crystals is small and therefore sensitive for extinction contrast, there is no average intensity decrease in the soft tissue regions of the samples (\fref{fig_2}). This implies that valuable information is encoded in low spatial frequencies within the angular acceptance, which are now made visible by the Bragg demagnifier through the enhanced PB-PCI. 
\\

\begin{figure}[t!]
\begin{center}
\includegraphics[scale=1.0]{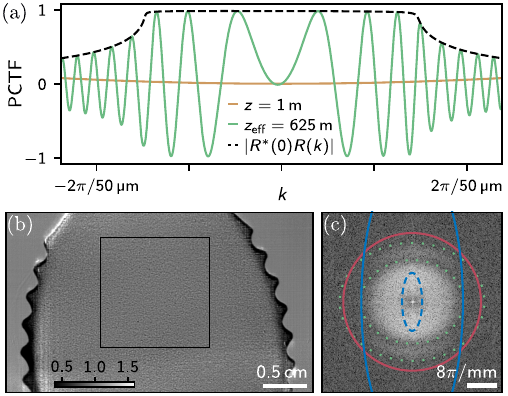}
\caption{
    (a) Phase contrast transfer function (see \eref{eq_2}) of weak phase objects at moderate resolution for conventional PB-PCI (brown) and using the Bragg demagnifier (green, black envelope) with a demagnification of $M=25$. 
    (b) Plastic spoon imaged with the Bragg demagnifier and (c) Fourier spectrum of the area marked in (b), plotted as $\log(|\tilde{I}(\mathbf{k})|)$. The spoon's rough surface acts as a weak phase noise object. The elliptic Thon rings originate from the roots of the PCTF with $z_{\text{eff},x}\approx\SI{380}{m}$ and $z_{\text{eff},y}\approx\SI{600}{m}$. The first two zero-crossings are indicated in green. Contrast degradation to $1/e$ at $z_x = \SI{0.57}{m}$ and $z_y = \SI{0.90}{m}$ due to the finite source size is shown by the solid blue line. Expected contrast degradation at a hypothetical physical propagation distance of $z_i=z_{\text{eff},i}$ is indicated by the dashed blue line. The red circle shows the $1/e$ contrast degradation by the optical microscope.
    } \label{fig_3}
    \end{center}
\end{figure}

In order to prove the presence of interference resulting from the increased effective propagation distance, we imaged a plastic spoon with a rough surface, which acts as a weak phase noise object. For weak phase objects, the contrast formation $I(\mathbf{r})=|g'(\mathbf{r})|^2$ reduces to a linear process, and the image intensity in Fourier space is given by $\tilde{I}(\mathbf{k}) \approx \delta(\mathbf{k}) - 2 \text{Im}[H^*(\mathbf{0})H(\mathbf{k})]\tilde{\varphi}(\mathbf{k})$~\cite{Lenz1958Mar,Kirkland2020Mar}. Here, $\tilde{\varphi}(\mathbf{k})$ denotes the phase shift in Fourier space of the sample exit wavefield. The absorption $\mu$ enters similarly via the real part but is omitted here for clarity. Insertion of the propagator yields
\begin{align}\label{eq_2}
    \tilde{I}(\mathbf{k}) \approx \delta(\mathbf{k}) + 2|R^*(\mathbf{0})R(\mathbf{k})|\sin\!\left(\frac{\mathbf{k}^2\mathbf{z}_\text{eff}}{2k_0} + \phi_R\!\right)\!\tilde{\varphi}(\mathbf{k}),
\end{align}
where $\phi_R$ is a slowly varying correction from the complex crystal reflectivity. The conversion of  phase information into image intensity defines the so-called phase contrast transfer function (PCTF). Analogously, replacing the sine by a cosine yields the amplitude contrast transfer function, which, however, only plays a minor role since $\mu$ is typically several orders of magnitude smaller than $\varphi$~\cite{momose1995phase}. As a side note, \eref{eq_2} emphasizes the need for large propagation distances for imaging low spatial frequencies. As an example, \frefadd{fig_3}{a} shows the free space PCTF for $z=\SI{1}{m}$ (brown). It generates almost no contrast. In contrast, a Bragg demagnifier with a demagnification factor of $M=25$ and a physical distance between crystal and detector of only \SI{1}{m} exhibits high values in the PCTF (green). The PCTF corresponds to conventional free space propagation with a very large distance of $z=\SI{625}{m}$, overlayed by the field amplitude reflection of the crystals (black).
In the Fourier spectrum of the plastic spoon so-called Thon rings become visible~(\frefadd{fig_3}{b,~c})~\cite{thon1966defokussierungsabhangigkeit}. They correspond to spatial frequencies where the PCTF is close to zero and are an additional evidence for the presence of interference effects. From the position of the Thon rings, we can extract a propagation distance of $z_{\text{eff},x}\approx\SI{380}{m}$ and $z_{\text{eff},y}\approx\SI{600}{m}$, in correspondence with the experimental settings.\\

Next, we consider the effect of source blur on the image degradation. The monochromator delivers an energy distribution and a corresponding angular distribution given by the size of the polychromatic source. The virtual source size is typically increased when the X-rays pass through asymmetric crystals~\cite{huang2012dispersive}. For our conditioner, we calculate a virtual spread by a factor of ${\sim}\,2$. Using matched crystals for the conditioner and demagnifier, the increase of the virtual source size can exactly be compensated~\cite{huang2012dispersive}. This means that the demagnified image experiences the source blur of its physical propagation distance $z$. Since the image was demagnified by a factor of $M$, the source blur is a factor of $M$ better compared to the image blur at a physical propagation distance of $M^2z$. This effect is demonstrated in \frefadd{fig_3}{c}. Assuming a Gaussian distribution for the source function, standard deviations of $\sigma_x=\SI{130}{\micro m}$ and $\sigma_y=\SI{30}{\micro m}$ in horizontal and vertical direction, respectively, have been extracted in preceding measurements. The resulting contrast degradation to $1/e$ at the physical propagation distance of $z_x=\SI{0.57}{m}$ and $z_y=\SI{0.90}{m}$ is depicted by the solid blue line. A hypothetical physical propagation distance of $z_i = z_{\text{eff},i}$ without demagnifier would suppress most of the image information, as shown by the dashed blue line. For completeness, we also indicate the $1/e$ degradation by the optical transfer function of the indirect detector system, see \cite{Spiecker}. The formation of additional Thon rings in vertical direction may further be suppressed since the spoon does not deliver a white noise power spectrum. \\

Instead of detecting the demagnified image with a high-resolution detector, the wavefield can also be re-magnified with a Bragg magnifier and detected by a highly-efficient large-area detector~\cite{Spiecker}. In this way, the Bragg demagnifier would enable imaging at moderate resolution with an even higher dose efficiency. Moreover, the shift-variance is exactly compensated when using the same crystals as for the demagnifier (see Appendix), allowing the image formation to be efficiently modeled in Fourier space.\\

In summary, we have presented a novel imaging technique to facilitate propagation-based X-ray phase contrast imaging of large, centimeter-sized samples at moderate resolution. By demagnifying the X-ray wavefield behind the sample with a Bragg demagnifier, the effective propagation distance is strongly increased. Thereby, the image contrast is significantly enhanced compared to phase contrast imaging at the same physical propagation distance, as we verified experimentally. Simultaneously, image blur by the finite source size is reduced. For the illumination of the sample, a Bragg conditioner is used in front of the sample to enlarge the beam cross section of the millimeter-sized synchrotron beam, thereby also increasing the monochromatic transversal coherence. By magnifying the propagated wavefield with a Bragg magnifier onto an efficient large-area detector, the technique may find utility for dose-efficient applications, e.g. in medical diagnostics in conjunction with compact X-ray sources~\cite{huang1998laser, Bech2009Jan, gunther2020versatile,hornberger2021inverse} for the early detection of breast cancer or other diseases. In future, the mechanical stability has to be further improved before larger crystals and higher energies as well as the combination with an additional Bragg magnifier can be investigated.

\section{Acknowledgments:}
We acknowledge DESY (Hamburg, Germany), a member of the Helmholtz Association HGF, for the provision of experimental facilities. Parts of this research were carried out at P23, PETRA~III, in the frame of the KIT-DESY 'Hierarchical Imaging Karlsruhe' collaboration, and we would like to thank Carlos Sato Baraldi Dias, Mateusz Czyzycki and Dmitri Novikov for assistance at the beamline. We are grateful to Simon Merz, Lea Bornemann (both Miltenyi Biotec) and Quentin Martinez (Stuttgart State Museum of Natural History) for providing the mouse liver lobe and salamander specimen, respectively. Further, we thank Stefan Uhlemann and Tobias Hilvering for technical assistance as well as Stephen Doyle for language revision.
The research was funded by the German Federal Ministry of Education and Research via grant 05K2019 (HIGH-LIFE).

\bibliography{bibliography}

\begin{thebibliography}{68}%
\makeatletter
\providecommand \@ifxundefined [1]{%
 \@ifx{#1\undefined}
}%
\providecommand \@ifnum [1]{%
 \ifnum #1\expandafter \@firstoftwo
 \else \expandafter \@secondoftwo
 \fi
}%
\providecommand \@ifx [1]{%
 \ifx #1\expandafter \@firstoftwo
 \else \expandafter \@secondoftwo
 \fi
}%
\providecommand \natexlab [1]{#1}%
\providecommand \enquote  [1]{``#1''}%
\providecommand \bibnamefont  [1]{#1}%
\providecommand \bibfnamefont [1]{#1}%
\providecommand \citenamefont [1]{#1}%
\providecommand \href@noop [0]{\@secondoftwo}%
\providecommand \href [0]{\begingroup \@sanitize@url \@href}%
\providecommand \@href[1]{\@@startlink{#1}\@@href}%
\providecommand \@@href[1]{\endgroup#1\@@endlink}%
\providecommand \@sanitize@url [0]{\catcode `\\12\catcode `\$12\catcode
  `\&12\catcode `\#12\catcode `\^12\catcode `\_12\catcode `\%12\relax}%
\providecommand \@@startlink[1]{}%
\providecommand \@@endlink[0]{}%
\providecommand \url  [0]{\begingroup\@sanitize@url \@url }%
\providecommand \@url [1]{\endgroup\@href {#1}{\urlprefix }}%
\providecommand \urlprefix  [0]{URL }%
\providecommand \Eprint [0]{\href }%
\providecommand \doibase [0]{https://doi.org/}%
\providecommand \selectlanguage [0]{\@gobble}%
\providecommand \bibinfo  [0]{\@secondoftwo}%
\providecommand \bibfield  [0]{\@secondoftwo}%
\providecommand \translation [1]{[#1]}%
\providecommand \BibitemOpen [0]{}%
\providecommand \bibitemStop [0]{}%
\providecommand \bibitemNoStop [0]{.\EOS\space}%
\providecommand \EOS [0]{\spacefactor3000\relax}%
\providecommand \BibitemShut  [1]{\csname bibitem#1\endcsname}%
\let\auto@bib@innerbib\@empty
\bibitem [{\citenamefont {Salditt}\ \emph {et~al.}(2017)\citenamefont
  {Salditt}, \citenamefont {Aspelmeier},\ and\ \citenamefont
  {Aeffner}}]{Salditt2017Oct}%
  \BibitemOpen
  \bibfield  {author} {\bibinfo {author} {\bibfnamefont {T.}~\bibnamefont
  {Salditt}}, \bibinfo {author} {\bibfnamefont {T.}~\bibnamefont
  {Aspelmeier}},\ and\ \bibinfo {author} {\bibfnamefont {S.}~\bibnamefont
  {Aeffner}},\ }\href {https://doi.org/10.1515/9783110426694} {\emph {\bibinfo
  {title} {{Biomedical Imaging}}}}\ (\bibinfo  {publisher} {De Gruyter},\
  \bibinfo {address} {Berlin, Germany},\ \bibinfo {year} {2017})\ p.\ \bibinfo
  {pages} {244ff.}\BibitemShut {Stop}%
\bibitem [{\citenamefont {Endrizzi}(2018)}]{endrizzi2018x}%
  \BibitemOpen
  \bibfield  {author} {\bibinfo {author} {\bibfnamefont {M.}~\bibnamefont
  {Endrizzi}},\ }\bibfield  {title} {\bibinfo {title} {{X}-ray phase-contrast
  imaging},\ }\href
  {https://doi.org/https://doi.org/10.1016/j.nima.2017.07.036} {\bibfield
  {journal} {\bibinfo  {journal} {Nuclear instruments and methods in physics
  research section A: Accelerators, spectrometers, detectors and associated
  equipment}\ }\textbf {\bibinfo {volume} {878}},\ \bibinfo {pages} {88}
  (\bibinfo {year} {2018})}\BibitemShut {NoStop}%
\bibitem [{\citenamefont {Chapman}\ and\ \citenamefont
  {Nugent}(2010)}]{Chapman2010Dec}%
  \BibitemOpen
  \bibfield  {author} {\bibinfo {author} {\bibfnamefont {H.~N.}\ \bibnamefont
  {Chapman}}\ and\ \bibinfo {author} {\bibfnamefont {K.~A.}\ \bibnamefont
  {Nugent}},\ }\bibfield  {title} {\bibinfo {title} {{Coherent lensless X-ray
  imaging}},\ }\href {https://doi.org/10.1038/nphoton.2010.240} {\bibfield
  {journal} {\bibinfo  {journal} {Nature Photonics}\ }\textbf {\bibinfo
  {volume} {4}},\ \bibinfo {pages} {833} (\bibinfo {year} {2010})}\BibitemShut
  {NoStop}%
\bibitem [{\citenamefont {Pfeiffer}\ \emph {et~al.}(2013)\citenamefont
  {Pfeiffer}, \citenamefont {Herzen}, \citenamefont {Willner}, \citenamefont
  {Chabior}, \citenamefont {Auweter}, \citenamefont {Reiser},\ and\
  \citenamefont {Bamberg}}]{Pfeiffer2013Sep}%
  \BibitemOpen
  \bibfield  {author} {\bibinfo {author} {\bibfnamefont {F.}~\bibnamefont
  {Pfeiffer}}, \bibinfo {author} {\bibfnamefont {J.}~\bibnamefont {Herzen}},
  \bibinfo {author} {\bibfnamefont {M.}~\bibnamefont {Willner}}, \bibinfo
  {author} {\bibfnamefont {M.}~\bibnamefont {Chabior}}, \bibinfo {author}
  {\bibfnamefont {S.}~\bibnamefont {Auweter}}, \bibinfo {author} {\bibfnamefont
  {M.}~\bibnamefont {Reiser}},\ and\ \bibinfo {author} {\bibfnamefont
  {F.}~\bibnamefont {Bamberg}},\ }\bibfield  {title} {\bibinfo {title}
  {{Grating-based {X}-ray phase contrast for biomedical imaging
  applications}},\ }\href {https://doi.org/10.1016/j.zemedi.2013.02.002}
  {\bibfield  {journal} {\bibinfo  {journal} {Z. Med. Phys.}\ }\textbf
  {\bibinfo {volume} {23}},\ \bibinfo {pages} {176} (\bibinfo {year}
  {2013})}\BibitemShut {NoStop}%
\bibitem [{\citenamefont {Lider}\ and\ \citenamefont
  {Kovalchuk}(2013)}]{Lider2013Nov}%
  \BibitemOpen
  \bibfield  {author} {\bibinfo {author} {\bibfnamefont {V.~V.}\ \bibnamefont
  {Lider}}\ and\ \bibinfo {author} {\bibfnamefont {M.~V.}\ \bibnamefont
  {Kovalchuk}},\ }\bibfield  {title} {\bibinfo {title} {{{X}-ray phase-contrast
  methods}},\ }\href {https://doi.org/10.1134/S1063774513050064} {\bibfield
  {journal} {\bibinfo  {journal} {Crystallogr. Rep.}\ }\textbf {\bibinfo
  {volume} {58}},\ \bibinfo {pages} {769} (\bibinfo {year} {2013})}\BibitemShut
  {NoStop}%
\bibitem [{\citenamefont {Tao}\ \emph {et~al.}(2021)\citenamefont {Tao},
  \citenamefont {He}, \citenamefont {Hao}, \citenamefont {Kuang},\ and\
  \citenamefont {Liu}}]{Tao2021Mar}%
  \BibitemOpen
  \bibfield  {author} {\bibinfo {author} {\bibfnamefont {S.}~\bibnamefont
  {Tao}}, \bibinfo {author} {\bibfnamefont {C.}~\bibnamefont {He}}, \bibinfo
  {author} {\bibfnamefont {X.}~\bibnamefont {Hao}}, \bibinfo {author}
  {\bibfnamefont {C.}~\bibnamefont {Kuang}},\ and\ \bibinfo {author}
  {\bibfnamefont {X.}~\bibnamefont {Liu}},\ }\bibfield  {title} {\bibinfo
  {title} {{Principles of Different {X}-ray Phase-Contrast Imaging: A
  Review}},\ }\href {https://doi.org/10.3390/app11072971} {\bibfield  {journal}
  {\bibinfo  {journal} {Appl. Sci.}\ }\textbf {\bibinfo {volume} {11}},\
  \bibinfo {pages} {2971} (\bibinfo {year} {2021})}\BibitemShut {NoStop}%
\bibitem [{\citenamefont {Mayo}\ \emph {et~al.}(2012)\citenamefont {Mayo},
  \citenamefont {Stevenson},\ and\ \citenamefont {Wilkins}}]{mayo2012line}%
  \BibitemOpen
  \bibfield  {author} {\bibinfo {author} {\bibfnamefont {S.~C.}\ \bibnamefont
  {Mayo}}, \bibinfo {author} {\bibfnamefont {A.~W.}\ \bibnamefont
  {Stevenson}},\ and\ \bibinfo {author} {\bibfnamefont {S.~W.}\ \bibnamefont
  {Wilkins}},\ }\bibfield  {title} {\bibinfo {title} {In-line phase-contrast
  {X}-ray imaging and tomography for materials science},\ }\href
  {https://doi.org/https://doi.org/10.3390/ma5050937} {\bibfield  {journal}
  {\bibinfo  {journal} {Materials}\ }\textbf {\bibinfo {volume} {5}},\ \bibinfo
  {pages} {937} (\bibinfo {year} {2012})}\BibitemShut {NoStop}%
\bibitem [{\citenamefont {Cotte}\ \emph {et~al.}(2019)\citenamefont {Cotte},
  \citenamefont {Autran}, \citenamefont {Berruyer}, \citenamefont {Dejoie},
  \citenamefont {Susini},\ and\ \citenamefont {Tafforeau}}]{cotte2019cultural}%
  \BibitemOpen
  \bibfield  {author} {\bibinfo {author} {\bibfnamefont {M.}~\bibnamefont
  {Cotte}}, \bibinfo {author} {\bibfnamefont {P.-O.}\ \bibnamefont {Autran}},
  \bibinfo {author} {\bibfnamefont {C.}~\bibnamefont {Berruyer}}, \bibinfo
  {author} {\bibfnamefont {C.}~\bibnamefont {Dejoie}}, \bibinfo {author}
  {\bibfnamefont {J.}~\bibnamefont {Susini}},\ and\ \bibinfo {author}
  {\bibfnamefont {P.}~\bibnamefont {Tafforeau}},\ }\bibfield  {title} {\bibinfo
  {title} {Cultural and natural heritage at the esrf: Looking back and to the
  future},\ }\href
  {https://doi.org/https://doi.org/10.1080/08940886.2019.1680213} {\bibfield
  {journal} {\bibinfo  {journal} {Synchrotron Radiation News}\ }\textbf
  {\bibinfo {volume} {32}},\ \bibinfo {pages} {34} (\bibinfo {year}
  {2019})}\BibitemShut {NoStop}%
\bibitem [{\citenamefont {Betz}\ \emph {et~al.}(2007)\citenamefont {Betz},
  \citenamefont {Wegst}, \citenamefont {Weide}, \citenamefont {Heethoff},
  \citenamefont {Helfen}, \citenamefont {Lee},\ and\ \citenamefont
  {Cloetens}}]{Betz2007Jul}%
  \BibitemOpen
  \bibfield  {author} {\bibinfo {author} {\bibfnamefont {O.}~\bibnamefont
  {Betz}}, \bibinfo {author} {\bibfnamefont {U.}~\bibnamefont {Wegst}},
  \bibinfo {author} {\bibfnamefont {D.}~\bibnamefont {Weide}}, \bibinfo
  {author} {\bibfnamefont {M.}~\bibnamefont {Heethoff}}, \bibinfo {author}
  {\bibfnamefont {L.}~\bibnamefont {Helfen}}, \bibinfo {author} {\bibfnamefont
  {W.-K.}\ \bibnamefont {Lee}},\ and\ \bibinfo {author} {\bibfnamefont
  {P.}~\bibnamefont {Cloetens}},\ }\bibfield  {title} {\bibinfo {title}
  {{Imaging applications of synchrotron {X}-ray phase-contrast microtomography
  in biological morphology and biomaterials science. I. General aspects of the
  technique and its advantages in the analysis of millimetre-sized arthropod
  structure}},\ }\href {https://doi.org/10.1111/j.1365-2818.2007.01785.x}
  {\bibfield  {journal} {\bibinfo  {journal} {J. Microsc.}\ }\textbf {\bibinfo
  {volume} {227}},\ \bibinfo {pages} {51} (\bibinfo {year} {2007})}\BibitemShut
  {NoStop}%
\bibitem [{\citenamefont {Kastner}\ \emph {et~al.}(2017)\citenamefont
  {Kastner}, \citenamefont {Heinzl}, \citenamefont {Plank}, \citenamefont
  {Salaberger}, \citenamefont {Gusenbauer},\ and\ \citenamefont
  {Senck}}]{kastner2017new}%
  \BibitemOpen
  \bibfield  {author} {\bibinfo {author} {\bibfnamefont {J.}~\bibnamefont
  {Kastner}}, \bibinfo {author} {\bibfnamefont {C.}~\bibnamefont {Heinzl}},
  \bibinfo {author} {\bibfnamefont {B.}~\bibnamefont {Plank}}, \bibinfo
  {author} {\bibfnamefont {D.}~\bibnamefont {Salaberger}}, \bibinfo {author}
  {\bibfnamefont {C.}~\bibnamefont {Gusenbauer}},\ and\ \bibinfo {author}
  {\bibfnamefont {S.}~\bibnamefont {Senck}},\ }\bibfield  {title} {\bibinfo
  {title} {New {X}-ray computed tomography methods for research and industry},\
  }in\ \href
  {https://www.ndt.net/events/iCT2017/app/content/Paper/99_Kastner_Rev1.pdf}
  {\emph {\bibinfo {booktitle} {7th Conference on Industrial Computed
  Tomography (iCT2017)}}}\ (\bibinfo {year} {2017})\BibitemShut {NoStop}%
\bibitem [{\citenamefont {Lewis}(2004)}]{lewis2004medical}%
  \BibitemOpen
  \bibfield  {author} {\bibinfo {author} {\bibfnamefont {R.~A.}\ \bibnamefont
  {Lewis}},\ }\bibfield  {title} {\bibinfo {title} {Medical phase contrast
  {X}-ray imaging: current status and future prospects},\ }\href
  {https://doi.org/10.1088/0031-9155/49/16/005} {\bibfield  {journal} {\bibinfo
   {journal} {Physics in medicine \& biology}\ }\textbf {\bibinfo {volume}
  {49}},\ \bibinfo {pages} {3573} (\bibinfo {year} {2004})}\BibitemShut
  {NoStop}%
\bibitem [{\citenamefont {Bravin}\ \emph {et~al.}(2012)\citenamefont {Bravin},
  \citenamefont {Coan},\ and\ \citenamefont {Suortti}}]{Bravin2012Dec}%
  \BibitemOpen
  \bibfield  {author} {\bibinfo {author} {\bibfnamefont {A.}~\bibnamefont
  {Bravin}}, \bibinfo {author} {\bibfnamefont {P.}~\bibnamefont {Coan}},\ and\
  \bibinfo {author} {\bibfnamefont {P.}~\bibnamefont {Suortti}},\ }\bibfield
  {title} {\bibinfo {title} {{{X}-ray phase-contrast imaging: from pre-clinical
  applications towards clinics}},\ }\href
  {https://doi.org/10.1088/0031-9155/58/1/R1} {\bibfield  {journal} {\bibinfo
  {journal} {Phys. Med. Biol.}\ }\textbf {\bibinfo {volume} {58}},\ \bibinfo
  {pages} {R1} (\bibinfo {year} {2012})}\BibitemShut {NoStop}%
\bibitem [{\citenamefont {Taba}\ \emph {et~al.}(2018)\citenamefont {Taba},
  \citenamefont {Gureyev}, \citenamefont {Alakhras}, \citenamefont {Lewis},
  \citenamefont {Lockie},\ and\ \citenamefont {Brennan}}]{Taba2018May}%
  \BibitemOpen
  \bibfield  {author} {\bibinfo {author} {\bibfnamefont {S.~T.}\ \bibnamefont
  {Taba}}, \bibinfo {author} {\bibfnamefont {T.~E.}\ \bibnamefont {Gureyev}},
  \bibinfo {author} {\bibfnamefont {M.}~\bibnamefont {Alakhras}}, \bibinfo
  {author} {\bibfnamefont {S.}~\bibnamefont {Lewis}}, \bibinfo {author}
  {\bibfnamefont {D.}~\bibnamefont {Lockie}},\ and\ \bibinfo {author}
  {\bibfnamefont {P.~C.}\ \bibnamefont {Brennan}},\ }\bibfield  {title}
  {\bibinfo {title} {{{X}-ray Phase-Contrast Technology in Breast Imaging:
  Principles, Options, and Clinical Application}},\ }\href
  {https://doi.org/10.2214/AJR.17.19179} {\bibfield  {journal} {\bibinfo
  {journal} {American Journal of Roentgenology}\ }\textbf {\bibinfo {volume}
  {211}},\ \bibinfo {pages} {133} (\bibinfo {year} {2018})}\BibitemShut
  {NoStop}%
\bibitem [{\citenamefont {Davis}\ \emph
  {et~al.}(1995{\natexlab{a}})\citenamefont {Davis}, \citenamefont {Gao},
  \citenamefont {Gureyev}, \citenamefont {Stevenson},\ and\ \citenamefont
  {Wilkins}}]{davis1995phase}%
  \BibitemOpen
  \bibfield  {author} {\bibinfo {author} {\bibfnamefont {T.~J.}\ \bibnamefont
  {Davis}}, \bibinfo {author} {\bibfnamefont {D.}~\bibnamefont {Gao}}, \bibinfo
  {author} {\bibfnamefont {T.}~\bibnamefont {Gureyev}}, \bibinfo {author}
  {\bibfnamefont {A.}~\bibnamefont {Stevenson}},\ and\ \bibinfo {author}
  {\bibfnamefont {S.}~\bibnamefont {Wilkins}},\ }\bibfield  {title} {\bibinfo
  {title} {Phase-contrast imaging of weakly absorbing materials using hard
  {X}-rays},\ }\href {https://doi.org/https://doi.org/10.1038/373595a0}
  {\bibfield  {journal} {\bibinfo  {journal} {Nature}\ }\textbf {\bibinfo
  {volume} {373}},\ \bibinfo {pages} {595} (\bibinfo {year}
  {1995}{\natexlab{a}})}\BibitemShut {NoStop}%
\bibitem [{\citenamefont {Chapman}\ \emph {et~al.}(1997)\citenamefont
  {Chapman}, \citenamefont {Thomlinson}, \citenamefont {Johnston},
  \citenamefont {Washburn}, \citenamefont {Pisano}, \citenamefont {Gm{\"u}r},
  \citenamefont {Zhong}, \citenamefont {Menk}, \citenamefont {Arfelli},\ and\
  \citenamefont {Sayers}}]{chapman1997diffraction}%
  \BibitemOpen
  \bibfield  {author} {\bibinfo {author} {\bibfnamefont {D.}~\bibnamefont
  {Chapman}}, \bibinfo {author} {\bibfnamefont {W.}~\bibnamefont {Thomlinson}},
  \bibinfo {author} {\bibfnamefont {R.}~\bibnamefont {Johnston}}, \bibinfo
  {author} {\bibfnamefont {D.}~\bibnamefont {Washburn}}, \bibinfo {author}
  {\bibfnamefont {E.}~\bibnamefont {Pisano}}, \bibinfo {author} {\bibfnamefont
  {N.}~\bibnamefont {Gm{\"u}r}}, \bibinfo {author} {\bibfnamefont
  {Z.}~\bibnamefont {Zhong}}, \bibinfo {author} {\bibfnamefont
  {R.}~\bibnamefont {Menk}}, \bibinfo {author} {\bibfnamefont {F.}~\bibnamefont
  {Arfelli}},\ and\ \bibinfo {author} {\bibfnamefont {D.}~\bibnamefont
  {Sayers}},\ }\bibfield  {title} {\bibinfo {title} {Diffraction enhanced
  {X}-ray imaging},\ }\href {https://doi.org/10.1088/0031-9155/42/11/001}
  {\bibfield  {journal} {\bibinfo  {journal} {Physics in Medicine \& Biology}\
  }\textbf {\bibinfo {volume} {42}},\ \bibinfo {pages} {2015} (\bibinfo {year}
  {1997})}\BibitemShut {NoStop}%
\bibitem [{\citenamefont {David}\ \emph {et~al.}(2002)\citenamefont {David},
  \citenamefont {N{\ifmmode\ddot{o}\else\"{o}\fi}hammer}, \citenamefont
  {Solak},\ and\ \citenamefont {Ziegler}}]{David2002Oct}%
  \BibitemOpen
  \bibfield  {author} {\bibinfo {author} {\bibfnamefont {C.}~\bibnamefont
  {David}}, \bibinfo {author} {\bibfnamefont {B.}~\bibnamefont
  {N{\ifmmode\ddot{o}\else\"{o}\fi}hammer}}, \bibinfo {author} {\bibfnamefont
  {H.~H.}\ \bibnamefont {Solak}},\ and\ \bibinfo {author} {\bibfnamefont
  {E.}~\bibnamefont {Ziegler}},\ }\bibfield  {title} {\bibinfo {title}
  {{Differential {X}-ray phase contrast imaging using a shearing
  interferometer}},\ }\href {https://doi.org/10.1063/1.1516611} {\bibfield
  {journal} {\bibinfo  {journal} {Appl. Phys. Lett.}\ }\textbf {\bibinfo
  {volume} {81}},\ \bibinfo {pages} {3287} (\bibinfo {year}
  {2002})}\BibitemShut {NoStop}%
\bibitem [{\citenamefont {Momose}\ \emph {et~al.}(2003)\citenamefont {Momose},
  \citenamefont {Kawamoto}, \citenamefont {Koyama}, \citenamefont {Hamaishi},
  \citenamefont {Takai},\ and\ \citenamefont
  {Suzuki}}]{momose2003demonstration}%
  \BibitemOpen
  \bibfield  {author} {\bibinfo {author} {\bibfnamefont {A.}~\bibnamefont
  {Momose}}, \bibinfo {author} {\bibfnamefont {S.}~\bibnamefont {Kawamoto}},
  \bibinfo {author} {\bibfnamefont {I.}~\bibnamefont {Koyama}}, \bibinfo
  {author} {\bibfnamefont {Y.}~\bibnamefont {Hamaishi}}, \bibinfo {author}
  {\bibfnamefont {K.}~\bibnamefont {Takai}},\ and\ \bibinfo {author}
  {\bibfnamefont {Y.}~\bibnamefont {Suzuki}},\ }\bibfield  {title} {\bibinfo
  {title} {Demonstration of {X}-ray talbot interferometry},\ }\href
  {https://doi.org/10.1143/JJAP.42.L866} {\bibfield  {journal} {\bibinfo
  {journal} {Japanese Journal of Applied Physics}\ }\textbf {\bibinfo {volume}
  {42}},\ \bibinfo {pages} {L866} (\bibinfo {year} {2003})}\BibitemShut
  {NoStop}%
\bibitem [{\citenamefont {Morgan}\ \emph {et~al.}(2012)\citenamefont {Morgan},
  \citenamefont {Paganin},\ and\ \citenamefont {Siu}}]{Morgan2012Mar}%
  \BibitemOpen
  \bibfield  {author} {\bibinfo {author} {\bibfnamefont {K.~S.}\ \bibnamefont
  {Morgan}}, \bibinfo {author} {\bibfnamefont {D.~M.}\ \bibnamefont
  {Paganin}},\ and\ \bibinfo {author} {\bibfnamefont {K.~K.~W.}\ \bibnamefont
  {Siu}},\ }\bibfield  {title} {\bibinfo {title} {{{X}-ray phase imaging with a
  paper analyzer}},\ }\bibfield  {journal} {\bibinfo  {journal} {Appl. Phys.
  Lett.}\ }\textbf {\bibinfo {volume} {100}},\ \href
  {https://doi.org/10.1063/1.3694918} {10.1063/1.3694918} (\bibinfo {year}
  {2012})\BibitemShut {NoStop}%
\bibitem [{\citenamefont {Zdora}(2018)}]{Zdora2018Apr}%
  \BibitemOpen
  \bibfield  {author} {\bibinfo {author} {\bibfnamefont {M.-C.}\ \bibnamefont
  {Zdora}},\ }\bibfield  {title} {\bibinfo {title} {{State of the Art of
  {X}-ray Speckle-Based Phase-Contrast and Dark-Field Imaging}},\ }\href
  {https://doi.org/10.3390/jimaging4050060} {\bibfield  {journal} {\bibinfo
  {journal} {J. Imaging}\ }\textbf {\bibinfo {volume} {4}},\ \bibinfo {pages}
  {60} (\bibinfo {year} {2018})}\BibitemShut {NoStop}%
\bibitem [{\citenamefont {Snigirev}\ \emph {et~al.}(1995)\citenamefont
  {Snigirev}, \citenamefont {Snigireva}, \citenamefont {Kohn}, \citenamefont
  {Kuznetsov},\ and\ \citenamefont {Schelokov}}]{snigirev1995possibilities}%
  \BibitemOpen
  \bibfield  {author} {\bibinfo {author} {\bibfnamefont {A.}~\bibnamefont
  {Snigirev}}, \bibinfo {author} {\bibfnamefont {I.}~\bibnamefont {Snigireva}},
  \bibinfo {author} {\bibfnamefont {V.}~\bibnamefont {Kohn}}, \bibinfo {author}
  {\bibfnamefont {S.}~\bibnamefont {Kuznetsov}},\ and\ \bibinfo {author}
  {\bibfnamefont {I.}~\bibnamefont {Schelokov}},\ }\bibfield  {title} {\bibinfo
  {title} {On the possibilities of {X}-ray phase contrast microimaging by
  coherent high-energy synchrotron radiation},\ }\href
  {https://doi.org/https://doi.org/10.1063/1.1146073} {\bibfield  {journal}
  {\bibinfo  {journal} {Review of Scientific Instruments}\ }\textbf {\bibinfo
  {volume} {66}},\ \bibinfo {pages} {5486} (\bibinfo {year}
  {1995})}\BibitemShut {NoStop}%
\bibitem [{\citenamefont {Wilkins}\ \emph {et~al.}(1996)\citenamefont
  {Wilkins}, \citenamefont {Gureyev}, \citenamefont {Gao}, \citenamefont
  {Pogany},\ and\ \citenamefont {Stevenson}}]{Wilkins1996Nov}%
  \BibitemOpen
  \bibfield  {author} {\bibinfo {author} {\bibfnamefont {S.~W.}\ \bibnamefont
  {Wilkins}}, \bibinfo {author} {\bibfnamefont {T.~E.}\ \bibnamefont
  {Gureyev}}, \bibinfo {author} {\bibfnamefont {D.}~\bibnamefont {Gao}},
  \bibinfo {author} {\bibfnamefont {A.}~\bibnamefont {Pogany}},\ and\ \bibinfo
  {author} {\bibfnamefont {A.~W.}\ \bibnamefont {Stevenson}},\ }\bibfield
  {title} {\bibinfo {title} {{Phase-contrast imaging using polychromatic hard
  {X}-rays}},\ }\href {https://doi.org/10.1038/384335a0} {\bibfield  {journal}
  {\bibinfo  {journal} {Nature}\ }\textbf {\bibinfo {volume} {384}},\ \bibinfo
  {pages} {335} (\bibinfo {year} {1996})}\BibitemShut {NoStop}%
\bibitem [{\citenamefont {Cloetens}\ \emph {et~al.}(1999)\citenamefont
  {Cloetens}, \citenamefont {Ludwig}, \citenamefont {Baruchel}, \citenamefont
  {Van~Dyck}, \citenamefont {Van~Landuyt}, \citenamefont {Guigay},\ and\
  \citenamefont {Schlenker}}]{cloetens1999holotomography}%
  \BibitemOpen
  \bibfield  {author} {\bibinfo {author} {\bibfnamefont {P.}~\bibnamefont
  {Cloetens}}, \bibinfo {author} {\bibfnamefont {W.}~\bibnamefont {Ludwig}},
  \bibinfo {author} {\bibfnamefont {J.}~\bibnamefont {Baruchel}}, \bibinfo
  {author} {\bibfnamefont {D.}~\bibnamefont {Van~Dyck}}, \bibinfo {author}
  {\bibfnamefont {J.}~\bibnamefont {Van~Landuyt}}, \bibinfo {author}
  {\bibfnamefont {J.}~\bibnamefont {Guigay}},\ and\ \bibinfo {author}
  {\bibfnamefont {M.}~\bibnamefont {Schlenker}},\ }\bibfield  {title} {\bibinfo
  {title} {Holotomography: Quantitative phase tomography with micrometer
  resolution using hard synchrotron radiation {X} rays},\ }\href
  {https://doi.org/https://doi.org/10.1063/1.125225} {\bibfield  {journal}
  {\bibinfo  {journal} {Applied Physics Letters}\ }\textbf {\bibinfo {volume}
  {75}},\ \bibinfo {pages} {2912} (\bibinfo {year} {1999})}\BibitemShut
  {NoStop}%
\bibitem [{\citenamefont {Zamir}\ \emph {et~al.}(2016)\citenamefont {Zamir},
  \citenamefont {Arthurs}, \citenamefont {Hagen}, \citenamefont {Diemoz},
  \citenamefont {Brochard}, \citenamefont {Bravin}, \citenamefont {Sebire},\
  and\ \citenamefont {Olivo}}]{zamir2016x}%
  \BibitemOpen
  \bibfield  {author} {\bibinfo {author} {\bibfnamefont {A.}~\bibnamefont
  {Zamir}}, \bibinfo {author} {\bibfnamefont {O.~J.}\ \bibnamefont {Arthurs}},
  \bibinfo {author} {\bibfnamefont {C.~K.}\ \bibnamefont {Hagen}}, \bibinfo
  {author} {\bibfnamefont {P.~C.}\ \bibnamefont {Diemoz}}, \bibinfo {author}
  {\bibfnamefont {T.}~\bibnamefont {Brochard}}, \bibinfo {author}
  {\bibfnamefont {A.}~\bibnamefont {Bravin}}, \bibinfo {author} {\bibfnamefont
  {N.~J.}\ \bibnamefont {Sebire}},\ and\ \bibinfo {author} {\bibfnamefont
  {A.}~\bibnamefont {Olivo}},\ }\bibfield  {title} {\bibinfo {title} {{X}-ray
  phase contrast tomography; proof of principle for post-mortem imaging},\
  }\href {https://doi.org/https://doi.org/10.1259/bjr.201505650} {\bibfield
  {journal} {\bibinfo  {journal} {The British Journal of Radiology}\ }\textbf
  {\bibinfo {volume} {89}},\ \bibinfo {pages} {20150565} (\bibinfo {year}
  {2016})}\BibitemShut {NoStop}%
\bibitem [{\citenamefont {Wagner}\ \emph {et~al.}(2018)\citenamefont {Wagner},
  \citenamefont {Wuennemann}, \citenamefont {Pacil{\'e}}, \citenamefont
  {Albers}, \citenamefont {Arfelli}, \citenamefont {Dreossi}, \citenamefont
  {Biederer}, \citenamefont {Konietzke}, \citenamefont {Stiller}, \citenamefont
  {Wielp{\"u}tz} \emph {et~al.}}]{wagner2018towards}%
  \BibitemOpen
  \bibfield  {author} {\bibinfo {author} {\bibfnamefont {W.~L.}\ \bibnamefont
  {Wagner}}, \bibinfo {author} {\bibfnamefont {F.}~\bibnamefont {Wuennemann}},
  \bibinfo {author} {\bibfnamefont {S.}~\bibnamefont {Pacil{\'e}}}, \bibinfo
  {author} {\bibfnamefont {J.}~\bibnamefont {Albers}}, \bibinfo {author}
  {\bibfnamefont {F.}~\bibnamefont {Arfelli}}, \bibinfo {author} {\bibfnamefont
  {D.}~\bibnamefont {Dreossi}}, \bibinfo {author} {\bibfnamefont
  {J.}~\bibnamefont {Biederer}}, \bibinfo {author} {\bibfnamefont
  {P.}~\bibnamefont {Konietzke}}, \bibinfo {author} {\bibfnamefont
  {W.}~\bibnamefont {Stiller}}, \bibinfo {author} {\bibfnamefont {M.~O.}\
  \bibnamefont {Wielp{\"u}tz}}, \emph {et~al.},\ }\bibfield  {title} {\bibinfo
  {title} {Towards synchrotron phase-contrast lung imaging in patients - a
  proof-of-concept study on porcine lungs in a human-scale chest phantom},\
  }\href {https://doi.org/https://doi.org/10.1107/S1600577518013401} {\bibfield
   {journal} {\bibinfo  {journal} {{Journal of Synchrotron Radiation}}\
  }\textbf {\bibinfo {volume} {25}},\ \bibinfo {pages} {1827} (\bibinfo {year}
  {2018})}\BibitemShut {NoStop}%
\bibitem [{\citenamefont {Donnelley}\ \emph {et~al.}(2019)\citenamefont
  {Donnelley}, \citenamefont {Morgan}, \citenamefont {Gradl}, \citenamefont
  {Klein}, \citenamefont {Hausermann}, \citenamefont {Hall}, \citenamefont
  {Maksimenko},\ and\ \citenamefont {Parsons}}]{donnelley2019live}%
  \BibitemOpen
  \bibfield  {author} {\bibinfo {author} {\bibfnamefont {M.}~\bibnamefont
  {Donnelley}}, \bibinfo {author} {\bibfnamefont {K.~S.}\ \bibnamefont
  {Morgan}}, \bibinfo {author} {\bibfnamefont {R.}~\bibnamefont {Gradl}},
  \bibinfo {author} {\bibfnamefont {M.}~\bibnamefont {Klein}}, \bibinfo
  {author} {\bibfnamefont {D.}~\bibnamefont {Hausermann}}, \bibinfo {author}
  {\bibfnamefont {C.}~\bibnamefont {Hall}}, \bibinfo {author} {\bibfnamefont
  {A.}~\bibnamefont {Maksimenko}},\ and\ \bibinfo {author} {\bibfnamefont
  {D.~W.}\ \bibnamefont {Parsons}},\ }\bibfield  {title} {\bibinfo {title}
  {{Live-pig-airway surface imaging and whole-pig CT at the Australian
  Synchrotron Imaging and Medical Beamline}},\ }\href
  {https://doi.org/https://doi.org/10.1107/S1600577518014133} {\bibfield
  {journal} {\bibinfo  {journal} {Journal of Synchrotron Radiation}\ }\textbf
  {\bibinfo {volume} {26}},\ \bibinfo {pages} {175} (\bibinfo {year}
  {2019})}\BibitemShut {NoStop}%
\bibitem [{\citenamefont {Walsh}\ \emph {et~al.}(2021)\citenamefont {Walsh},
  \citenamefont {Tafforeau}, \citenamefont {Wagner}, \citenamefont {Jafree},
  \citenamefont {Bellier}, \citenamefont {Werlein}, \citenamefont {K{\"u}hnel},
  \citenamefont {Boller}, \citenamefont {Walker-Samuel}, \citenamefont
  {Robertus} \emph {et~al.}}]{walsh2021imaging}%
  \BibitemOpen
  \bibfield  {author} {\bibinfo {author} {\bibfnamefont {C.}~\bibnamefont
  {Walsh}}, \bibinfo {author} {\bibfnamefont {P.}~\bibnamefont {Tafforeau}},
  \bibinfo {author} {\bibfnamefont {W.}~\bibnamefont {Wagner}}, \bibinfo
  {author} {\bibfnamefont {D.}~\bibnamefont {Jafree}}, \bibinfo {author}
  {\bibfnamefont {A.}~\bibnamefont {Bellier}}, \bibinfo {author} {\bibfnamefont
  {C.}~\bibnamefont {Werlein}}, \bibinfo {author} {\bibfnamefont
  {M.}~\bibnamefont {K{\"u}hnel}}, \bibinfo {author} {\bibfnamefont
  {E.}~\bibnamefont {Boller}}, \bibinfo {author} {\bibfnamefont
  {S.}~\bibnamefont {Walker-Samuel}}, \bibinfo {author} {\bibfnamefont
  {J.}~\bibnamefont {Robertus}}, \emph {et~al.},\ }\bibfield  {title} {\bibinfo
  {title} {Imaging intact human organs with local resolution of cellular
  structures using hierarchical phase-contrast tomography},\ }\href
  {https://doi.org/https://doi.org/10.1038/s41592-021-01317-x} {\bibfield
  {journal} {\bibinfo  {journal} {Nature Methods}\ }\textbf {\bibinfo {volume}
  {18}},\ \bibinfo {pages} {1532} (\bibinfo {year} {2021})}\BibitemShut
  {NoStop}%
\bibitem [{\citenamefont {Tavakoli}\ \emph {et~al.}(2021)\citenamefont
  {Tavakoli}, \citenamefont {Cuccione}, \citenamefont {Dumot}, \citenamefont
  {Cormode}, \citenamefont {Wiart}, \citenamefont {Elleaume},\ and\
  \citenamefont {Brun}}]{tavakoli2021tracking}%
  \BibitemOpen
  \bibfield  {author} {\bibinfo {author} {\bibfnamefont {C.}~\bibnamefont
  {Tavakoli}}, \bibinfo {author} {\bibfnamefont {E.}~\bibnamefont {Cuccione}},
  \bibinfo {author} {\bibfnamefont {C.}~\bibnamefont {Dumot}}, \bibinfo
  {author} {\bibfnamefont {D.}~\bibnamefont {Cormode}}, \bibinfo {author}
  {\bibfnamefont {M.}~\bibnamefont {Wiart}}, \bibinfo {author} {\bibfnamefont
  {H.}~\bibnamefont {Elleaume}},\ and\ \bibinfo {author} {\bibfnamefont
  {E.}~\bibnamefont {Brun}},\ }\bibfield  {title} {\bibinfo {title} {Tracking
  cells in the brain of small animals using synchrotron multi-spectral phase
  contrast imaging},\ }in\ \href
  {https://doi.org/https://doi.org/10.1117/12.2580841} {\emph {\bibinfo
  {booktitle} {Medical Imaging 2021: Physics of Medical Imaging}}},\ Vol.\
  \bibinfo {volume} {11595}\ (\bibinfo {organization} {SPIE},\ \bibinfo {year}
  {2021})\ pp.\ \bibinfo {pages} {1218--1224}\BibitemShut {NoStop}%
\bibitem [{\citenamefont {Arhatari}\ \emph {et~al.}(2021)\citenamefont
  {Arhatari}, \citenamefont {Stevenson}, \citenamefont {Abbey}, \citenamefont
  {Nesterets}, \citenamefont {Maksimenko}, \citenamefont {Hall}, \citenamefont
  {Thompson}, \citenamefont {Mayo}, \citenamefont {Fiala}, \citenamefont
  {Quiney} \emph {et~al.}}]{arhatari2021x}%
  \BibitemOpen
  \bibfield  {author} {\bibinfo {author} {\bibfnamefont {B.~D.}\ \bibnamefont
  {Arhatari}}, \bibinfo {author} {\bibfnamefont {A.~W.}\ \bibnamefont
  {Stevenson}}, \bibinfo {author} {\bibfnamefont {B.}~\bibnamefont {Abbey}},
  \bibinfo {author} {\bibfnamefont {Y.~I.}\ \bibnamefont {Nesterets}}, \bibinfo
  {author} {\bibfnamefont {A.}~\bibnamefont {Maksimenko}}, \bibinfo {author}
  {\bibfnamefont {C.~J.}\ \bibnamefont {Hall}}, \bibinfo {author}
  {\bibfnamefont {D.}~\bibnamefont {Thompson}}, \bibinfo {author}
  {\bibfnamefont {S.~C.}\ \bibnamefont {Mayo}}, \bibinfo {author}
  {\bibfnamefont {T.}~\bibnamefont {Fiala}}, \bibinfo {author} {\bibfnamefont
  {H.~M.}\ \bibnamefont {Quiney}}, \emph {et~al.},\ }\bibfield  {title}
  {\bibinfo {title} {{{X}-ray phase-contrast computed tomography for soft
  tissue imaging at the imaging and medical beamline ({IMBL}) of the Australian
  Synchrotron}},\ }\href {https://doi.org/https://doi.org/10.3390/app11094120}
  {\bibfield  {journal} {\bibinfo  {journal} {Applied Sciences}\ }\textbf
  {\bibinfo {volume} {11}},\ \bibinfo {pages} {4120} (\bibinfo {year}
  {2021})}\BibitemShut {NoStop}%
\bibitem [{\citenamefont {Cotte}\ \emph {et~al.}(2022)\citenamefont {Cotte},
  \citenamefont {Dollman}, \citenamefont {Fernandez}, \citenamefont {Gonzalez},
  \citenamefont {Vanmeert}, \citenamefont {Monico}, \citenamefont {Dejoie},
  \citenamefont {Burghammer}, \citenamefont {Huder}, \citenamefont {Fisher}
  \emph {et~al.}}]{cotte2022new}%
  \BibitemOpen
  \bibfield  {author} {\bibinfo {author} {\bibfnamefont {M.}~\bibnamefont
  {Cotte}}, \bibinfo {author} {\bibfnamefont {K.}~\bibnamefont {Dollman}},
  \bibinfo {author} {\bibfnamefont {V.}~\bibnamefont {Fernandez}}, \bibinfo
  {author} {\bibfnamefont {V.}~\bibnamefont {Gonzalez}}, \bibinfo {author}
  {\bibfnamefont {F.}~\bibnamefont {Vanmeert}}, \bibinfo {author}
  {\bibfnamefont {L.}~\bibnamefont {Monico}}, \bibinfo {author} {\bibfnamefont
  {C.}~\bibnamefont {Dejoie}}, \bibinfo {author} {\bibfnamefont
  {M.}~\bibnamefont {Burghammer}}, \bibinfo {author} {\bibfnamefont
  {L.}~\bibnamefont {Huder}}, \bibinfo {author} {\bibfnamefont
  {S.}~\bibnamefont {Fisher}}, \emph {et~al.},\ }\bibfield  {title} {\bibinfo
  {title} {{New Opportunities Offered by the ESRF to the Cultural and Natural
  Heritage Communities}},\ }\href
  {https://doi.org/https://doi.org/10.1080/08940886.2022.2135958} {\bibfield
  {journal} {\bibinfo  {journal} {Synchrotron Radiation News}\ }\textbf
  {\bibinfo {volume} {35}},\ \bibinfo {pages} {3} (\bibinfo {year}
  {2022})}\BibitemShut {NoStop}%
\bibitem [{\citenamefont {H{\"a}ggmark}\ \emph {et~al.}(2023)\citenamefont
  {H{\"a}ggmark}, \citenamefont {Shaker}, \citenamefont {Nyr{\'e}n},
  \citenamefont {Al-Amiry}, \citenamefont {Abadi}, \citenamefont {P.~Segars},
  \citenamefont {Samei},\ and\ \citenamefont {M.~Hertz}}]{haggmark2023phase}%
  \BibitemOpen
  \bibfield  {author} {\bibinfo {author} {\bibfnamefont {I.}~\bibnamefont
  {H{\"a}ggmark}}, \bibinfo {author} {\bibfnamefont {K.}~\bibnamefont
  {Shaker}}, \bibinfo {author} {\bibfnamefont {S.}~\bibnamefont {Nyr{\'e}n}},
  \bibinfo {author} {\bibfnamefont {B.}~\bibnamefont {Al-Amiry}}, \bibinfo
  {author} {\bibfnamefont {E.}~\bibnamefont {Abadi}}, \bibinfo {author}
  {\bibfnamefont {W.}~\bibnamefont {P.~Segars}}, \bibinfo {author}
  {\bibfnamefont {E.}~\bibnamefont {Samei}},\ and\ \bibinfo {author}
  {\bibfnamefont {H.}~\bibnamefont {M.~Hertz}},\ }\bibfield  {title} {\bibinfo
  {title} {Phase-contrast virtual chest radiography},\ }\href
  {https://doi.org/https://doi.org/10.1073/pnas.2210214120} {\bibfield
  {journal} {\bibinfo  {journal} {Proceedings of the National Academy of
  Sciences}\ }\textbf {\bibinfo {volume} {120}},\ \bibinfo {pages}
  {e2210214120} (\bibinfo {year} {2023})}\BibitemShut {NoStop}%
\bibitem [{\citenamefont {Cianciosi}\ \emph {et~al.}(2021)\citenamefont
  {Cianciosi}, \citenamefont {Buisson}, \citenamefont {Tafforeau},\ and\
  \citenamefont {Van~Vaerenbergh}}]{Cianciosi2021Oct}%
  \BibitemOpen
  \bibfield  {author} {\bibinfo {author} {\bibfnamefont {F.}~\bibnamefont
  {Cianciosi}}, \bibinfo {author} {\bibfnamefont {A.-L.}\ \bibnamefont
  {Buisson}}, \bibinfo {author} {\bibfnamefont {P.}~\bibnamefont {Tafforeau}},\
  and\ \bibinfo {author} {\bibfnamefont {P.}~\bibnamefont {Van~Vaerenbergh}},\
  }\href {https://doi.org/10.18429/JACoW-MEDSI2020-MOIO02} {\emph {\bibinfo
  {title} {BM18, the New ESRF-EBS Beamline for Hierarchical Phase-Contrast
  Tomography}}}\ (\bibinfo  {publisher} {JACOW Publishing, Geneva,
  Switzerland},\ \bibinfo {year} {2021})\BibitemShut {NoStop}%
\bibitem [{\citenamefont {Lang}\ \emph {et~al.}(2023)\citenamefont {Lang},
  \citenamefont {Saeidnezhad}, \citenamefont {Dremel}, \citenamefont {Weller},
  \citenamefont {Diez}, \citenamefont {Stock}, \citenamefont {Sauer},
  \citenamefont {Cianciosi}, \citenamefont {Jarnias}, \citenamefont
  {Tafforeau},\ and\ \citenamefont {Zabler}}]{Lang2023Mar}%
  \BibitemOpen
  \bibfield  {author} {\bibinfo {author} {\bibfnamefont {T.}~\bibnamefont
  {Lang}}, \bibinfo {author} {\bibfnamefont {N.}~\bibnamefont {Saeidnezhad}},
  \bibinfo {author} {\bibfnamefont {K.}~\bibnamefont {Dremel}}, \bibinfo
  {author} {\bibfnamefont {D.}~\bibnamefont {Weller}}, \bibinfo {author}
  {\bibfnamefont {M.}~\bibnamefont {Diez}}, \bibinfo {author} {\bibfnamefont
  {A.~M.}\ \bibnamefont {Stock}}, \bibinfo {author} {\bibfnamefont
  {T.}~\bibnamefont {Sauer}}, \bibinfo {author} {\bibfnamefont
  {F.}~\bibnamefont {Cianciosi}}, \bibinfo {author} {\bibfnamefont
  {C.}~\bibnamefont {Jarnias}}, \bibinfo {author} {\bibfnamefont
  {P.}~\bibnamefont {Tafforeau}},\ and\ \bibinfo {author} {\bibfnamefont
  {S.}~\bibnamefont {Zabler}},\ }\bibfield  {title} {\bibinfo {title}
  {{Multiscale Phase-Contrast Tomography at BM18}},\ }\href
  {https://doi.org/10.58286/27746} {\bibfield  {journal} {\bibinfo  {journal}
  {e-Journal of Nondestructive Testing}\ }\textbf {\bibinfo {volume} {28}}
  (\bibinfo {year} {2023})}\BibitemShut {NoStop}%
\bibitem [{\citenamefont {Spiecker}\ \emph {et~al.}(2023)\citenamefont
  {Spiecker}, \citenamefont {Pfeiffer}, \citenamefont {Biswal}, \citenamefont
  {Shcherbinin}, \citenamefont {Spiecker}, \citenamefont {Hessdorfer},
  \citenamefont {Hurst}, \citenamefont {Zharov}, \citenamefont {Bellucci},
  \citenamefont {Farago}, \citenamefont {Zuber}, \citenamefont {Herz},
  \citenamefont {Cecilia}, \citenamefont {Czyzycki}, \citenamefont {Dias},
  \citenamefont {Novikov}, \citenamefont {Krogmann}, \citenamefont {Hamann},
  \citenamefont {van~de Kamp},\ and\ \citenamefont {Baumbach}}]{Spiecker}%
  \BibitemOpen
  \bibfield  {author} {\bibinfo {author} {\bibfnamefont {R.}~\bibnamefont
  {Spiecker}}, \bibinfo {author} {\bibfnamefont {P.}~\bibnamefont {Pfeiffer}},
  \bibinfo {author} {\bibfnamefont {A.}~\bibnamefont {Biswal}}, \bibinfo
  {author} {\bibfnamefont {M.}~\bibnamefont {Shcherbinin}}, \bibinfo {author}
  {\bibfnamefont {M.}~\bibnamefont {Spiecker}}, \bibinfo {author}
  {\bibfnamefont {H.}~\bibnamefont {Hessdorfer}}, \bibinfo {author}
  {\bibfnamefont {M.}~\bibnamefont {Hurst}}, \bibinfo {author} {\bibfnamefont
  {Y.}~\bibnamefont {Zharov}}, \bibinfo {author} {\bibfnamefont
  {V.}~\bibnamefont {Bellucci}}, \bibinfo {author} {\bibfnamefont
  {T.}~\bibnamefont {Farago}}, \bibinfo {author} {\bibfnamefont
  {M.}~\bibnamefont {Zuber}}, \bibinfo {author} {\bibfnamefont
  {A.}~\bibnamefont {Herz}}, \bibinfo {author} {\bibfnamefont {A.}~\bibnamefont
  {Cecilia}}, \bibinfo {author} {\bibfnamefont {M.}~\bibnamefont {Czyzycki}},
  \bibinfo {author} {\bibfnamefont {C.}~\bibnamefont {Dias}}, \bibinfo {author}
  {\bibfnamefont {D.}~\bibnamefont {Novikov}}, \bibinfo {author} {\bibfnamefont
  {L.}~\bibnamefont {Krogmann}}, \bibinfo {author} {\bibfnamefont
  {E.}~\bibnamefont {Hamann}}, \bibinfo {author} {\bibfnamefont
  {T.}~\bibnamefont {van~de Kamp}},\ and\ \bibinfo {author} {\bibfnamefont
  {T.}~\bibnamefont {Baumbach}},\ }\bibfield  {title} {\bibinfo {title}
  {{Dose-efficient in vivo X-ray phase contrast imaging at micrometer
  resolution}},\ }\bibfield  {journal} {\bibinfo  {journal} {Optica}\ }\href
  {https://doi.org/10.1364/OPTICA.500978} {10.1364/OPTICA.500978} (\bibinfo
  {year} {2023})\BibitemShut {NoStop}%
\bibitem [{\citenamefont {Cowley}(1978)}]{Cowley1978Jan}%
  \BibitemOpen
  \bibfield  {author} {\bibinfo {author} {\bibfnamefont {J.~M.}\ \bibnamefont
  {Cowley}},\ }\bibfield  {title} {\bibinfo {title} {{Electron
  Microdiffraction}},\ }in\ \href
  {https://doi.org/10.1016/S0065-2539(08)60410-2} {\emph {\bibinfo {booktitle}
  {{Advances in Electronics and Electron Physics}}}},\ Vol.~\bibinfo {volume}
  {46}\ (\bibinfo  {publisher} {Academic Press},\ \bibinfo {address}
  {Cambridge, MA, USA},\ \bibinfo {year} {1978})\ pp.\ \bibinfo {pages}
  {1--53}\BibitemShut {NoStop}%
\bibitem [{\citenamefont {Boettinger}\ \emph {et~al.}(1979)\citenamefont
  {Boettinger}, \citenamefont {Burdette},\ and\ \citenamefont
  {Kuriyama}}]{boettinger1979x}%
  \BibitemOpen
  \bibfield  {author} {\bibinfo {author} {\bibfnamefont {W.~J.}\ \bibnamefont
  {Boettinger}}, \bibinfo {author} {\bibfnamefont {H.~E.}\ \bibnamefont
  {Burdette}},\ and\ \bibinfo {author} {\bibfnamefont {M.}~\bibnamefont
  {Kuriyama}},\ }\bibfield  {title} {\bibinfo {title} {{X}-ray magnifier},\
  }\href {https://doi.org/https://doi.org/10.1063/1.1135662} {\bibfield
  {journal} {\bibinfo  {journal} {Review of Scientific Instruments}\ }\textbf
  {\bibinfo {volume} {50}},\ \bibinfo {pages} {26} (\bibinfo {year}
  {1979})}\BibitemShut {NoStop}%
\bibitem [{\citenamefont {Kobayashi}\ \emph {et~al.}(2001)\citenamefont
  {Kobayashi}, \citenamefont {Izumi}, \citenamefont {Kimura}, \citenamefont
  {Kimura}, \citenamefont {Ibuki}, \citenamefont {Yokoyama}, \citenamefont
  {Tsusaka}, \citenamefont {Kagoshima},\ and\ \citenamefont
  {Matsui}}]{kobayashi2001x}%
  \BibitemOpen
  \bibfield  {author} {\bibinfo {author} {\bibfnamefont {K.}~\bibnamefont
  {Kobayashi}}, \bibinfo {author} {\bibfnamefont {K.}~\bibnamefont {Izumi}},
  \bibinfo {author} {\bibfnamefont {H.}~\bibnamefont {Kimura}}, \bibinfo
  {author} {\bibfnamefont {S.}~\bibnamefont {Kimura}}, \bibinfo {author}
  {\bibfnamefont {T.}~\bibnamefont {Ibuki}}, \bibinfo {author} {\bibfnamefont
  {Y.}~\bibnamefont {Yokoyama}}, \bibinfo {author} {\bibfnamefont
  {Y.}~\bibnamefont {Tsusaka}}, \bibinfo {author} {\bibfnamefont
  {Y.}~\bibnamefont {Kagoshima}},\ and\ \bibinfo {author} {\bibfnamefont
  {J.}~\bibnamefont {Matsui}},\ }\bibfield  {title} {\bibinfo {title} {{X}-ray
  phase-contrast imaging with submicron resolution by using extremely
  asymmetric {B}ragg diffractions},\ }\href
  {https://doi.org/https://doi.org/10.1063/1.1337621} {\bibfield  {journal}
  {\bibinfo  {journal} {Applied Physics Letters}\ }\textbf {\bibinfo {volume}
  {78}},\ \bibinfo {pages} {132} (\bibinfo {year} {2001})}\BibitemShut
  {NoStop}%
\bibitem [{\citenamefont {Stampanoni}\ \emph {et~al.}(2002)\citenamefont
  {Stampanoni}, \citenamefont {Borchert}, \citenamefont {Abela},\ and\
  \citenamefont {R{\"u}egsegger}}]{stampanoni2002bragg}%
  \BibitemOpen
  \bibfield  {author} {\bibinfo {author} {\bibfnamefont {M.}~\bibnamefont
  {Stampanoni}}, \bibinfo {author} {\bibfnamefont {G.}~\bibnamefont
  {Borchert}}, \bibinfo {author} {\bibfnamefont {R.}~\bibnamefont {Abela}},\
  and\ \bibinfo {author} {\bibfnamefont {P.}~\bibnamefont {R{\"u}egsegger}},\
  }\bibfield  {title} {\bibinfo {title} {{B}ragg magnifier: A detector for
  submicrometer {X}-ray computer tomography},\ }\href
  {https://doi.org/https://doi.org/10.1063/1.1520722} {\bibfield  {journal}
  {\bibinfo  {journal} {Journal of Applied Physics}\ }\textbf {\bibinfo
  {volume} {92}},\ \bibinfo {pages} {7630} (\bibinfo {year}
  {2002})}\BibitemShut {NoStop}%
\bibitem [{\citenamefont {Modregger}\ \emph {et~al.}(2006)\citenamefont
  {Modregger}, \citenamefont {L{\"u}bbert}, \citenamefont {Sch{\"a}fer},\ and\
  \citenamefont {K{\"o}hler}}]{modregger2006magnified}%
  \BibitemOpen
  \bibfield  {author} {\bibinfo {author} {\bibfnamefont {P.}~\bibnamefont
  {Modregger}}, \bibinfo {author} {\bibfnamefont {D.}~\bibnamefont
  {L{\"u}bbert}}, \bibinfo {author} {\bibfnamefont {P.}~\bibnamefont
  {Sch{\"a}fer}},\ and\ \bibinfo {author} {\bibfnamefont {R.}~\bibnamefont
  {K{\"o}hler}},\ }\bibfield  {title} {\bibinfo {title} {Magnified {X}-ray
  phase imaging using asymmetric {B}ragg reflection: Experiment and theory},\
  }\href {https://doi.org/https://doi.org/10.1103/PhysRevB.74.054107}
  {\bibfield  {journal} {\bibinfo  {journal} {Physical Review B}\ }\textbf
  {\bibinfo {volume} {74}},\ \bibinfo {pages} {054107} (\bibinfo {year}
  {2006})}\BibitemShut {NoStop}%
\bibitem [{\citenamefont {Watanabe}\ \emph {et~al.}(1999)\citenamefont
  {Watanabe}, \citenamefont {Suzuki}, \citenamefont {Higashi},\ and\
  \citenamefont {Sakabe}}]{watanabe1999rotated}%
  \BibitemOpen
  \bibfield  {author} {\bibinfo {author} {\bibfnamefont {N.}~\bibnamefont
  {Watanabe}}, \bibinfo {author} {\bibfnamefont {M.}~\bibnamefont {Suzuki}},
  \bibinfo {author} {\bibfnamefont {Y.}~\bibnamefont {Higashi}},\ and\ \bibinfo
  {author} {\bibfnamefont {N.}~\bibnamefont {Sakabe}},\ }\bibfield  {title}
  {\bibinfo {title} {Rotated-inclined focusing monochromator with simultaneous
  tuning of asymmetry factor and radius of curvature over a wide wavelength
  range},\ }\href {https://doi.org/https://doi.org/10.1107/S0909049599000229}
  {\bibfield  {journal} {\bibinfo  {journal} {Journal of Synchrotron
  Radiation}\ }\textbf {\bibinfo {volume} {6}},\ \bibinfo {pages} {64}
  (\bibinfo {year} {1999})}\BibitemShut {NoStop}%
\bibitem [{\citenamefont {Tsusaka}\ \emph {et~al.}(2000)\citenamefont
  {Tsusaka}, \citenamefont {Yokoyama}, \citenamefont {Takeda}, \citenamefont
  {Urakawa}, \citenamefont {Kagoshima}, \citenamefont {Matsui}, \citenamefont
  {Kimura}, \citenamefont {Kimura}, \citenamefont {Kobayashi},\ and\
  \citenamefont {Izumi}}]{tsusaka2000formation}%
  \BibitemOpen
  \bibfield  {author} {\bibinfo {author} {\bibfnamefont {Y.}~\bibnamefont
  {Tsusaka}}, \bibinfo {author} {\bibfnamefont {K.}~\bibnamefont {Yokoyama}},
  \bibinfo {author} {\bibfnamefont {S.}~\bibnamefont {Takeda}}, \bibinfo
  {author} {\bibfnamefont {M.}~\bibnamefont {Urakawa}}, \bibinfo {author}
  {\bibfnamefont {Y.}~\bibnamefont {Kagoshima}}, \bibinfo {author}
  {\bibfnamefont {J.}~\bibnamefont {Matsui}}, \bibinfo {author} {\bibfnamefont
  {S.}~\bibnamefont {Kimura}}, \bibinfo {author} {\bibfnamefont
  {H.}~\bibnamefont {Kimura}}, \bibinfo {author} {\bibfnamefont
  {K.}~\bibnamefont {Kobayashi}},\ and\ \bibinfo {author} {\bibfnamefont
  {K.}~\bibnamefont {Izumi}},\ }\bibfield  {title} {\bibinfo {title} {Formation
  of parallel {X}-ray microbeam and its application},\ }\href
  {https://doi.org/10.1143/JJAP.39.L635} {\bibfield  {journal} {\bibinfo
  {journal} {Japanese Journal of Applied Physics}\ }\textbf {\bibinfo {volume}
  {39}},\ \bibinfo {pages} {L635} (\bibinfo {year} {2000})}\BibitemShut
  {NoStop}%
\bibitem [{\citenamefont {Hirano}\ and\ \citenamefont
  {Takahashi}(2013)}]{hirano2013applications}%
  \BibitemOpen
  \bibfield  {author} {\bibinfo {author} {\bibfnamefont {K.}~\bibnamefont
  {Hirano}}\ and\ \bibinfo {author} {\bibfnamefont {Y.}~\bibnamefont
  {Takahashi}},\ }\bibfield  {title} {\bibinfo {title} {Applications of {X}-ray
  magnifier and demagnifier to angle-resolved {X}-ray computed tomography},\
  }in\ \href {https://doi.org/10.1088/1742-6596/425/19/192004} {\emph {\bibinfo
  {booktitle} {Journal of Physics: Conference Series}}},\ Vol.\ \bibinfo
  {volume} {425}\ (\bibinfo {organization} {IOP Publishing},\ \bibinfo {year}
  {2013})\ p.\ \bibinfo {pages} {192004}\BibitemShut {NoStop}%
\bibitem [{\citenamefont {Authier}(2004)}]{authier2004dynamical}%
  \BibitemOpen
  \bibfield  {author} {\bibinfo {author} {\bibfnamefont {A.}~\bibnamefont
  {Authier}},\ }\href@noop {} {\emph {\bibinfo {title} {Dynamical theory of
  {X}-ray diffraction}}},\ Vol.~\bibinfo {volume} {11}\ (\bibinfo  {publisher}
  {Oxford University Press Inc.},\ \bibinfo {address} {New York, United
  States},\ \bibinfo {year} {2004})\ p.\ \bibinfo {pages} {189ff.}\BibitemShut
  {Stop}%
\bibitem [{\citenamefont {Abbe}(1873)}]{abbe1873beitrage}%
  \BibitemOpen
  \bibfield  {author} {\bibinfo {author} {\bibfnamefont {E.}~\bibnamefont
  {Abbe}},\ }\bibfield  {title} {\bibinfo {title} {{Beitr{\"a}ge zur Theorie
  des Mikroskops und der mikroskopischen Wahrnehmung}},\ }\href
  {https://doi.org/https://doi.org/10.1007/BF02956173} {\bibfield  {journal}
  {\bibinfo  {journal} {Archiv f{\"u}r mikroskopische Anatomie}\ }\textbf
  {\bibinfo {volume} {9}},\ \bibinfo {pages} {413} (\bibinfo {year}
  {1873})}\BibitemShut {NoStop}%
\bibitem [{\citenamefont {Martinson}\ \emph {et~al.}(2015)\citenamefont
  {Martinson}, \citenamefont {Samadi}, \citenamefont {Bassey}, \citenamefont
  {Gomez},\ and\ \citenamefont {Chapman}}]{Martinson2015May}%
  \BibitemOpen
  \bibfield  {author} {\bibinfo {author} {\bibfnamefont {M.}~\bibnamefont
  {Martinson}}, \bibinfo {author} {\bibfnamefont {N.}~\bibnamefont {Samadi}},
  \bibinfo {author} {\bibfnamefont {B.}~\bibnamefont {Bassey}}, \bibinfo
  {author} {\bibfnamefont {A.}~\bibnamefont {Gomez}},\ and\ \bibinfo {author}
  {\bibfnamefont {D.}~\bibnamefont {Chapman}},\ }\bibfield  {title} {\bibinfo
  {title} {{Phase-preserving beam expander for biomedical {X}-ray imaging}},\
  }\href {https://doi.org/10.1107/S1600577515004695} {\bibfield  {journal}
  {\bibinfo  {journal} {J. Synchrotron Radiat.}\ }\textbf {\bibinfo {volume}
  {22}},\ \bibinfo {pages} {801} (\bibinfo {year} {2015})},\ \Eprint
  {https://arxiv.org/abs/25931100} {25931100} \BibitemShut {NoStop}%
\bibitem [{\citenamefont {M{\'a}rkus}\ \emph {et~al.}(2018)\citenamefont
  {M{\'a}rkus}, \citenamefont {Greving}, \citenamefont {Kornemann},
  \citenamefont {Storm}, \citenamefont {Beckmann}, \citenamefont {Mohr},\ and\
  \citenamefont {Last}}]{markus2018optimizing}%
  \BibitemOpen
  \bibfield  {author} {\bibinfo {author} {\bibfnamefont {O.}~\bibnamefont
  {M{\'a}rkus}}, \bibinfo {author} {\bibfnamefont {I.}~\bibnamefont {Greving}},
  \bibinfo {author} {\bibfnamefont {E.}~\bibnamefont {Kornemann}}, \bibinfo
  {author} {\bibfnamefont {M.}~\bibnamefont {Storm}}, \bibinfo {author}
  {\bibfnamefont {F.}~\bibnamefont {Beckmann}}, \bibinfo {author}
  {\bibfnamefont {J.}~\bibnamefont {Mohr}},\ and\ \bibinfo {author}
  {\bibfnamefont {A.}~\bibnamefont {Last}},\ }\bibfield  {title} {\bibinfo
  {title} {Optimizing illumination for full field imaging at high brilliance
  hard {X}-ray synchrotron sources},\ }\href
  {https://doi.org/https://doi.org/10.1364/OE.26.030435} {\bibfield  {journal}
  {\bibinfo  {journal} {Optics express}\ }\textbf {\bibinfo {volume} {26}},\
  \bibinfo {pages} {30435} (\bibinfo {year} {2018})}\BibitemShut {NoStop}%
\bibitem [{\citenamefont {Reinhard}\ \emph {et~al.}(2021)\citenamefont
  {Reinhard}, \citenamefont {Drakopoulos}, \citenamefont {Ahmed}, \citenamefont
  {Deyhle}, \citenamefont {James}, \citenamefont {Charlesworth}, \citenamefont
  {Burt}, \citenamefont {Sutter}, \citenamefont {Alexander}, \citenamefont
  {Garland} \emph {et~al.}}]{reinhard2021beamline}%
  \BibitemOpen
  \bibfield  {author} {\bibinfo {author} {\bibfnamefont {C.}~\bibnamefont
  {Reinhard}}, \bibinfo {author} {\bibfnamefont {M.}~\bibnamefont
  {Drakopoulos}}, \bibinfo {author} {\bibfnamefont {S.~I.}\ \bibnamefont
  {Ahmed}}, \bibinfo {author} {\bibfnamefont {H.}~\bibnamefont {Deyhle}},
  \bibinfo {author} {\bibfnamefont {A.}~\bibnamefont {James}}, \bibinfo
  {author} {\bibfnamefont {C.~M.}\ \bibnamefont {Charlesworth}}, \bibinfo
  {author} {\bibfnamefont {M.}~\bibnamefont {Burt}}, \bibinfo {author}
  {\bibfnamefont {J.}~\bibnamefont {Sutter}}, \bibinfo {author} {\bibfnamefont
  {S.}~\bibnamefont {Alexander}}, \bibinfo {author} {\bibfnamefont
  {P.}~\bibnamefont {Garland}}, \emph {et~al.},\ }\bibfield  {title} {\bibinfo
  {title} {{Beamline K11 DIAD: A new instrument for dual imaging and
  diffraction at Diamond Light Source}},\ }\href
  {https://doi.org/https://doi.org/10.1107/S1600577521009875} {\bibfield
  {journal} {\bibinfo  {journal} {Journal of Synchrotron Radiation}\ }\textbf
  {\bibinfo {volume} {28}},\ \bibinfo {pages} {1985} (\bibinfo {year}
  {2021})}\BibitemShut {NoStop}%
\bibitem [{\citenamefont {Matsushita}\ and\ \citenamefont
  {Hashizume}(1983)}]{Koch1983}%
  \BibitemOpen
  \bibfield  {author} {\bibinfo {author} {\bibfnamefont {T.}~\bibnamefont
  {Matsushita}}\ and\ \bibinfo {author} {\bibfnamefont {H.}~\bibnamefont
  {Hashizume}},\ }\href
  {https://books.google.de/books/about/Handbook_on_Synchrotron_Radiation.html?id=xF1SrgEACAAJ&redir_esc=y}
  {\emph {\bibinfo {title} {{Handbook on Synchrotron Radiation}}}},\ edited by\
  \bibinfo {editor} {\bibfnamefont {E.}~\bibnamefont {Koch}},\ Vol.~\bibinfo
  {volume} {1}\ (\bibinfo  {publisher} {North Holland Publishing Company},\
  \bibinfo {year} {1983})\ Chap.~\bibinfo {chapter} {4}\BibitemShut {NoStop}%
\bibitem [{\citenamefont {Christensen}\ \emph {et~al.}(1994)\citenamefont
  {Christensen}, \citenamefont {Hornstrup}, \citenamefont {Frederiksen},
  \citenamefont {Abdali}, \citenamefont {Grundsoe}, \citenamefont {Schnopper},
  \citenamefont {Lewis}, \citenamefont {Hall},\ and\ \citenamefont
  {Borozdin}}]{christensen1994expanded}%
  \BibitemOpen
  \bibfield  {author} {\bibinfo {author} {\bibfnamefont {F.~E.}\ \bibnamefont
  {Christensen}}, \bibinfo {author} {\bibfnamefont {A.}~\bibnamefont
  {Hornstrup}}, \bibinfo {author} {\bibfnamefont {P.~K.}\ \bibnamefont
  {Frederiksen}}, \bibinfo {author} {\bibfnamefont {S.}~\bibnamefont {Abdali}},
  \bibinfo {author} {\bibfnamefont {P.}~\bibnamefont {Grundsoe}}, \bibinfo
  {author} {\bibfnamefont {H.~W.}\ \bibnamefont {Schnopper}}, \bibinfo {author}
  {\bibfnamefont {R.~A.}\ \bibnamefont {Lewis}}, \bibinfo {author}
  {\bibfnamefont {C.~J.}\ \bibnamefont {Hall}},\ and\ \bibinfo {author}
  {\bibfnamefont {K.~N.}\ \bibnamefont {Borozdin}},\ }\bibfield  {title}
  {\bibinfo {title} {{Expanded beam {X}-ray optics calibration facility at the
  Daresbury Synchrotron}},\ }in\ \href
  {https://doi.org/https://doi.org/10.1117/12.167225} {\emph {\bibinfo
  {booktitle} {Multilayer and Grazing Incidence {X}-ray/EUV Optics II}}},\
  Vol.\ \bibinfo {volume} {2011}\ (\bibinfo {organization} {SPIE},\ \bibinfo
  {year} {1994})\ pp.\ \bibinfo {pages} {540--548}\BibitemShut {NoStop}%
\bibitem [{\citenamefont {Kamezawa}\ \emph {et~al.}(2023)\citenamefont
  {Kamezawa}, \citenamefont {Hyodo}, \citenamefont {Tokunaga}, \citenamefont
  {Tsukada},\ and\ \citenamefont {Matushita}}]{Kamezawa2023Sep}%
  \BibitemOpen
  \bibfield  {author} {\bibinfo {author} {\bibfnamefont {C.}~\bibnamefont
  {Kamezawa}}, \bibinfo {author} {\bibfnamefont {K.}~\bibnamefont {Hyodo}},
  \bibinfo {author} {\bibfnamefont {C.}~\bibnamefont {Tokunaga}}, \bibinfo
  {author} {\bibfnamefont {T.}~\bibnamefont {Tsukada}},\ and\ \bibinfo {author}
  {\bibfnamefont {S.}~\bibnamefont {Matushita}},\ }\bibfield  {title} {\bibinfo
  {title} {{Large-view {X}-ray imaging for medical applications using the
  world{'}s only vertically polarized synchrotron radiation beam and a single
  asymmetric Si crystal}},\ }\href {https://doi.org/10.1088/1361-6560/acf640}
  {\bibfield  {journal} {\bibinfo  {journal} {Phys. Med. Biol.}\ }\textbf
  {\bibinfo {volume} {68}},\ \bibinfo {pages} {195010} (\bibinfo {year}
  {2023})}\BibitemShut {NoStop}%
\bibitem [{\citenamefont {Davis}\ \emph
  {et~al.}(1995{\natexlab{b}})\citenamefont {Davis}, \citenamefont {Gureyev},
  \citenamefont {Gao}, \citenamefont {Stevenson},\ and\ \citenamefont
  {Wilkins}}]{Davis1995Apr}%
  \BibitemOpen
  \bibfield  {author} {\bibinfo {author} {\bibfnamefont {T.~J.}\ \bibnamefont
  {Davis}}, \bibinfo {author} {\bibfnamefont {T.~E.}\ \bibnamefont {Gureyev}},
  \bibinfo {author} {\bibfnamefont {D.}~\bibnamefont {Gao}}, \bibinfo {author}
  {\bibfnamefont {A.~W.}\ \bibnamefont {Stevenson}},\ and\ \bibinfo {author}
  {\bibfnamefont {S.~W.}\ \bibnamefont {Wilkins}},\ }\bibfield  {title}
  {\bibinfo {title} {{{X}-ray Image Contrast from a Simple Phase Object}},\
  }\href {https://doi.org/10.1103/PhysRevLett.74.3173} {\bibfield  {journal}
  {\bibinfo  {journal} {Phys. Rev. Lett.}\ }\textbf {\bibinfo {volume} {74}},\
  \bibinfo {pages} {3173} (\bibinfo {year} {1995}{\natexlab{b}})}\BibitemShut
  {NoStop}%
\bibitem [{\citenamefont {Shastri}(2004)}]{Shastri2004Mar}%
  \BibitemOpen
  \bibfield  {author} {\bibinfo {author} {\bibfnamefont {S.~D.}\ \bibnamefont
  {Shastri}},\ }\bibfield  {title} {\bibinfo {title} {{Combining flat crystals,
  bent crystals and compound refractive lenses for high-energy X-ray optics}},\
  }\href {https://doi.org/10.1107/S0909049503023586} {\bibfield  {journal}
  {\bibinfo  {journal} {Journal of Synchrotron Radiatiation}\ }\textbf
  {\bibinfo {volume} {11}},\ \bibinfo {pages} {150} (\bibinfo {year}
  {2004})}\BibitemShut {NoStop}%
\bibitem [{\citenamefont {Bravin}(2003)}]{bravin2003exploiting}%
  \BibitemOpen
  \bibfield  {author} {\bibinfo {author} {\bibfnamefont {A.}~\bibnamefont
  {Bravin}},\ }\bibfield  {title} {\bibinfo {title} {Exploiting the {X}-ray
  refraction contrast with an analyser: the state of the art},\ }\href
  {https://doi.org/10.1088/0022-3727/36/10A/306} {\bibfield  {journal}
  {\bibinfo  {journal} {Journal of Physics D: Applied Physics}\ }\textbf
  {\bibinfo {volume} {36}},\ \bibinfo {pages} {A24} (\bibinfo {year}
  {2003})}\BibitemShut {NoStop}%
\bibitem [{\citenamefont {Lenz}\ and\ \citenamefont
  {Scheffels}(1958)}]{Lenz1958Mar}%
  \BibitemOpen
  \bibfield  {author} {\bibinfo {author} {\bibfnamefont {F.}~\bibnamefont
  {Lenz}}\ and\ \bibinfo {author} {\bibfnamefont {W.}~\bibnamefont
  {Scheffels}},\ }\bibfield  {title} {\bibinfo {title} {{Das Zusammenwirken von
  Phasen- und Amplitudenkontrast in der elektronenmikroskopischen Abbildung}},\
  }\href {https://doi.org/10.1515/zna-1958-0309} {\bibfield  {journal}
  {\bibinfo  {journal} {Zeitschrift f{\ifmmode\ddot{u}\else\"{u}\fi}r
  Naturforschung A}\ }\textbf {\bibinfo {volume} {13}},\ \bibinfo {pages} {226}
  (\bibinfo {year} {1958})}\BibitemShut {NoStop}%
\bibitem [{\citenamefont {Kirkland}(2020)}]{Kirkland2020Mar}%
  \BibitemOpen
  \bibfield  {author} {\bibinfo {author} {\bibfnamefont {E.~J.}\ \bibnamefont
  {Kirkland}},\ }\bibfield  {title} {\bibinfo {title} {{Some Image
  Approximations}},\ }in\ \href {https://doi.org/10.1007/978-3-030-33260-0_3}
  {\emph {\bibinfo {booktitle} {{Advanced Computing in Electron Microscopy}}}}\
  (\bibinfo  {publisher} {Springer},\ \bibinfo {address} {Cham, Switzerland},\
  \bibinfo {year} {2020})\ pp.\ \bibinfo {pages} {37--80}\BibitemShut {NoStop}%
\bibitem [{\citenamefont {Momose}\ and\ \citenamefont
  {Fukuda}(1995)}]{momose1995phase}%
  \BibitemOpen
  \bibfield  {author} {\bibinfo {author} {\bibfnamefont {A.}~\bibnamefont
  {Momose}}\ and\ \bibinfo {author} {\bibfnamefont {J.}~\bibnamefont
  {Fukuda}},\ }\bibfield  {title} {\bibinfo {title} {Phase-contrast radiographs
  of nonstained rat cerebellar specimen},\ }\href
  {https://doi.org/https://doi.org/10.1118/1.597472} {\bibfield  {journal}
  {\bibinfo  {journal} {Medical Physics}\ }\textbf {\bibinfo {volume} {22}},\
  \bibinfo {pages} {375} (\bibinfo {year} {1995})}\BibitemShut {NoStop}%
\bibitem [{\citenamefont {Thon}(1966)}]{thon1966defokussierungsabhangigkeit}%
  \BibitemOpen
  \bibfield  {author} {\bibinfo {author} {\bibfnamefont {F.}~\bibnamefont
  {Thon}},\ }\bibfield  {title} {\bibinfo {title} {{Zur
  Defokussierungsabh{\"a}ngigkeit des Phasenkontrastes bei der
  elektronenmikroskopischen Abbildung}},\ }\href
  {https://doi.org/https://doi.org/10.1515/zna-1966-0417} {\bibfield  {journal}
  {\bibinfo  {journal} {Zeitschrift f{\"u}r Naturforschung A}\ }\textbf
  {\bibinfo {volume} {21}},\ \bibinfo {pages} {476} (\bibinfo {year}
  {1966})}\BibitemShut {NoStop}%
\bibitem [{\citenamefont {Huang}\ \emph {et~al.}(2012)\citenamefont {Huang},
  \citenamefont {Macrander}, \citenamefont {Honnicke}, \citenamefont {Cai},\
  and\ \citenamefont {Fernandez}}]{huang2012dispersive}%
  \BibitemOpen
  \bibfield  {author} {\bibinfo {author} {\bibfnamefont {X.}~\bibnamefont
  {Huang}}, \bibinfo {author} {\bibfnamefont {A.}~\bibnamefont {Macrander}},
  \bibinfo {author} {\bibfnamefont {M.}~\bibnamefont {Honnicke}}, \bibinfo
  {author} {\bibfnamefont {Y.}~\bibnamefont {Cai}},\ and\ \bibinfo {author}
  {\bibfnamefont {P.}~\bibnamefont {Fernandez}},\ }\bibfield  {title} {\bibinfo
  {title} {Dispersive spread of virtual sources by asymmetric {X}-ray
  monochromators},\ }\href
  {https://doi.org/https://doi.org/10.1107/S0021889812003366} {\bibfield
  {journal} {\bibinfo  {journal} {Journal of Applied Crystallography}\ }\textbf
  {\bibinfo {volume} {45}},\ \bibinfo {pages} {255} (\bibinfo {year}
  {2012})}\BibitemShut {NoStop}%
\bibitem [{\citenamefont {Huang}\ and\ \citenamefont
  {Ruth}(1998)}]{huang1998laser}%
  \BibitemOpen
  \bibfield  {author} {\bibinfo {author} {\bibfnamefont {Z.}~\bibnamefont
  {Huang}}\ and\ \bibinfo {author} {\bibfnamefont {R.~D.}\ \bibnamefont
  {Ruth}},\ }\bibfield  {title} {\bibinfo {title} {Laser-electron storage
  ring},\ }\href {https://doi.org/https://doi.org/10.1103/PhysRevLett.80.976}
  {\bibfield  {journal} {\bibinfo  {journal} {Physical Review Letters}\
  }\textbf {\bibinfo {volume} {80}},\ \bibinfo {pages} {976} (\bibinfo {year}
  {1998})}\BibitemShut {NoStop}%
\bibitem [{\citenamefont {Bech}\ \emph {et~al.}(2009)\citenamefont {Bech},
  \citenamefont {Bunk}, \citenamefont {David}, \citenamefont {Ruth},
  \citenamefont {Rifkin}, \citenamefont {Loewen}, \citenamefont
  {Feidenhans'l},\ and\ \citenamefont {Pfeiffer}}]{Bech2009Jan}%
  \BibitemOpen
  \bibfield  {author} {\bibinfo {author} {\bibfnamefont {M.}~\bibnamefont
  {Bech}}, \bibinfo {author} {\bibfnamefont {O.}~\bibnamefont {Bunk}}, \bibinfo
  {author} {\bibfnamefont {C.}~\bibnamefont {David}}, \bibinfo {author}
  {\bibfnamefont {R.}~\bibnamefont {Ruth}}, \bibinfo {author} {\bibfnamefont
  {J.}~\bibnamefont {Rifkin}}, \bibinfo {author} {\bibfnamefont
  {R.}~\bibnamefont {Loewen}}, \bibinfo {author} {\bibfnamefont
  {R.}~\bibnamefont {Feidenhans'l}},\ and\ \bibinfo {author} {\bibfnamefont
  {F.}~\bibnamefont {Pfeiffer}},\ }\bibfield  {title} {\bibinfo {title} {{Hard
  X-ray phase-contrast imaging with the Compact Light Source based on inverse
  Compton X-rays}},\ }\href {https://doi.org/10.1107/S090904950803464X}
  {\bibfield  {journal} {\bibinfo  {journal} {Journal of Synchrotron
  Radiation}\ }\textbf {\bibinfo {volume} {16}},\ \bibinfo {pages} {43}
  (\bibinfo {year} {2009})}\BibitemShut {NoStop}%
\bibitem [{\citenamefont {G{\"u}nther}\ \emph {et~al.}(2020)\citenamefont
  {G{\"u}nther}, \citenamefont {Gradl}, \citenamefont {Jud}, \citenamefont
  {Eggl}, \citenamefont {Huang}, \citenamefont {Kulpe}, \citenamefont
  {Achterhold}, \citenamefont {Gleich}, \citenamefont {Dierolf},\ and\
  \citenamefont {Pfeiffer}}]{gunther2020versatile}%
  \BibitemOpen
  \bibfield  {author} {\bibinfo {author} {\bibfnamefont {B.}~\bibnamefont
  {G{\"u}nther}}, \bibinfo {author} {\bibfnamefont {R.}~\bibnamefont {Gradl}},
  \bibinfo {author} {\bibfnamefont {C.}~\bibnamefont {Jud}}, \bibinfo {author}
  {\bibfnamefont {E.}~\bibnamefont {Eggl}}, \bibinfo {author} {\bibfnamefont
  {J.}~\bibnamefont {Huang}}, \bibinfo {author} {\bibfnamefont
  {S.}~\bibnamefont {Kulpe}}, \bibinfo {author} {\bibfnamefont
  {K.}~\bibnamefont {Achterhold}}, \bibinfo {author} {\bibfnamefont
  {B.}~\bibnamefont {Gleich}}, \bibinfo {author} {\bibfnamefont
  {M.}~\bibnamefont {Dierolf}},\ and\ \bibinfo {author} {\bibfnamefont
  {F.}~\bibnamefont {Pfeiffer}},\ }\bibfield  {title} {\bibinfo {title} {{The
  versatile {X}-ray beamline of the Munich Compact Light Source: design,
  instrumentation and applications}},\ }\href
  {https://doi.org/https://doi.org/10.1107/S1600577520008309} {\bibfield
  {journal} {\bibinfo  {journal} {Journal of Synchrotron Radiation}\ }\textbf
  {\bibinfo {volume} {27}},\ \bibinfo {pages} {1395} (\bibinfo {year}
  {2020})}\BibitemShut {NoStop}%
\bibitem [{\citenamefont {Hornberger}\ \emph {et~al.}(2021)\citenamefont
  {Hornberger}, \citenamefont {Kasahara}, \citenamefont {Ruth}, \citenamefont
  {Loewen},\ and\ \citenamefont {Khaydarov}}]{hornberger2021inverse}%
  \BibitemOpen
  \bibfield  {author} {\bibinfo {author} {\bibfnamefont {B.}~\bibnamefont
  {Hornberger}}, \bibinfo {author} {\bibfnamefont {J.}~\bibnamefont
  {Kasahara}}, \bibinfo {author} {\bibfnamefont {R.}~\bibnamefont {Ruth}},
  \bibinfo {author} {\bibfnamefont {R.}~\bibnamefont {Loewen}},\ and\ \bibinfo
  {author} {\bibfnamefont {J.}~\bibnamefont {Khaydarov}},\ }\bibfield  {title}
  {\bibinfo {title} {Inverse compton scattering {X}-ray source for research,
  industry and medical applications},\ }in\ \href
  {https://doi.org/https://doi.org/10.1117/12.2591977} {\emph {\bibinfo
  {booktitle} {International Conference on {X}-ray Lasers 2020}}},\ Vol.\
  \bibinfo {volume} {11886}\ (\bibinfo {organization} {SPIE},\ \bibinfo {year}
  {2021})\ pp.\ \bibinfo {pages} {51--60}\BibitemShut {NoStop}%
\bibitem [{\citenamefont {Modregger}\ \emph {et~al.}(2008)\citenamefont
  {Modregger}, \citenamefont {L{\ifmmode\ddot{u}\else\"{u}\fi}bbert},
  \citenamefont {Sch{\ifmmode\ddot{a}\else\"{a}\fi}fer}, \citenamefont
  {K{\ifmmode\ddot{o}\else\"{o}\fi}hler}, \citenamefont {Weitkamp},
  \citenamefont {Hanke},\ and\ \citenamefont {Baumbach}}]{Modregger2008Mar}%
  \BibitemOpen
  \bibfield  {author} {\bibinfo {author} {\bibfnamefont {P.}~\bibnamefont
  {Modregger}}, \bibinfo {author} {\bibfnamefont {D.}~\bibnamefont
  {L{\ifmmode\ddot{u}\else\"{u}\fi}bbert}}, \bibinfo {author} {\bibfnamefont
  {P.}~\bibnamefont {Sch{\ifmmode\ddot{a}\else\"{a}\fi}fer}}, \bibinfo {author}
  {\bibfnamefont {R.}~\bibnamefont {K{\ifmmode\ddot{o}\else\"{o}\fi}hler}},
  \bibinfo {author} {\bibfnamefont {T.}~\bibnamefont {Weitkamp}}, \bibinfo
  {author} {\bibfnamefont {M.}~\bibnamefont {Hanke}},\ and\ \bibinfo {author}
  {\bibfnamefont {T.}~\bibnamefont {Baumbach}},\ }\bibfield  {title} {\bibinfo
  {title} {{Fresnel diffraction in the case of an inclined image plane}},\
  }\href {https://doi.org/10.1364/OE.16.005141} {\bibfield  {journal} {\bibinfo
   {journal} {Opt. Express}\ }\textbf {\bibinfo {volume} {16}},\ \bibinfo
  {pages} {5141} (\bibinfo {year} {2008})}\BibitemShut {NoStop}%
\bibitem [{\citenamefont {Hriv{\v{n}}ak}\ \emph {et~al.}(2018)\citenamefont
  {Hriv{\v{n}}ak}, \citenamefont {Uli{\v{c}}n{\`y}},\ and\ \citenamefont
  {Vagovi{\v{c}}}}]{hrivvnak2018fast}%
  \BibitemOpen
  \bibfield  {author} {\bibinfo {author} {\bibfnamefont {S.}~\bibnamefont
  {Hriv{\v{n}}ak}}, \bibinfo {author} {\bibfnamefont {J.}~\bibnamefont
  {Uli{\v{c}}n{\`y}}},\ and\ \bibinfo {author} {\bibfnamefont {P.}~\bibnamefont
  {Vagovi{\v{c}}}},\ }\bibfield  {title} {\bibinfo {title} {Fast {F}resnel
  propagation through a set of inclined reflecting planes applicable for
  {X}-ray imaging},\ }\href
  {https://doi.org/https://doi.org/10.1364/OE.26.034569} {\bibfield  {journal}
  {\bibinfo  {journal} {Optics Express}\ }\textbf {\bibinfo {volume} {26}},\
  \bibinfo {pages} {34569} (\bibinfo {year} {2018})}\BibitemShut {NoStop}%
\bibitem [{\citenamefont {Vagovi{\v{c}}}\ \emph {et~al.}(2014)\citenamefont
  {Vagovi{\v{c}}}, \citenamefont {{\v{S}}v{\'e}da}, \citenamefont {Cecilia},
  \citenamefont {Hamann}, \citenamefont {Pelliccia}, \citenamefont {Gimenez},
  \citenamefont {Koryt{\'a}r}, \citenamefont {Pavlov}, \citenamefont
  {Z{\'a}pra{\v{z}}n{\`y}}, \citenamefont {Zuber} \emph
  {et~al.}}]{vagovivc2014x}%
  \BibitemOpen
  \bibfield  {author} {\bibinfo {author} {\bibfnamefont {P.}~\bibnamefont
  {Vagovi{\v{c}}}}, \bibinfo {author} {\bibfnamefont {L.}~\bibnamefont
  {{\v{S}}v{\'e}da}}, \bibinfo {author} {\bibfnamefont {A.}~\bibnamefont
  {Cecilia}}, \bibinfo {author} {\bibfnamefont {E.}~\bibnamefont {Hamann}},
  \bibinfo {author} {\bibfnamefont {D.}~\bibnamefont {Pelliccia}}, \bibinfo
  {author} {\bibfnamefont {E.}~\bibnamefont {Gimenez}}, \bibinfo {author}
  {\bibfnamefont {D.}~\bibnamefont {Koryt{\'a}r}}, \bibinfo {author}
  {\bibfnamefont {K.~M.}\ \bibnamefont {Pavlov}}, \bibinfo {author}
  {\bibfnamefont {Z.}~\bibnamefont {Z{\'a}pra{\v{z}}n{\`y}}}, \bibinfo {author}
  {\bibfnamefont {M.}~\bibnamefont {Zuber}}, \emph {et~al.},\ }\bibfield
  {title} {\bibinfo {title} {{X}-ray bragg magnifier microscope as a linear
  shift invariant imaging system: image formation and phase retrieval},\ }\href
  {https://doi.org/https://doi.org/10.1364/OE.22.021508} {\bibfield  {journal}
  {\bibinfo  {journal} {Optics Express}\ }\textbf {\bibinfo {volume} {22}},\
  \bibinfo {pages} {21508} (\bibinfo {year} {2014})}\BibitemShut {NoStop}%
\bibitem [{\citenamefont {Hriv{\v{n}}ak}\ \emph {et~al.}(2016)\citenamefont
  {Hriv{\v{n}}ak}, \citenamefont {Uli{\v{c}}n{\`y}}, \citenamefont
  {Mike{\v{s}}}, \citenamefont {Cecilia}, \citenamefont {Hamann}, \citenamefont
  {Baumbach}, \citenamefont {{\v{S}}v{\'e}da}, \citenamefont
  {Z{\'a}pra{\v{z}}n{\`y}}, \citenamefont {Koryt{\'a}r}, \citenamefont
  {Gimenez-Navarro} \emph {et~al.}}]{hrivvnak2016single}%
  \BibitemOpen
  \bibfield  {author} {\bibinfo {author} {\bibfnamefont {S.}~\bibnamefont
  {Hriv{\v{n}}ak}}, \bibinfo {author} {\bibfnamefont {J.}~\bibnamefont
  {Uli{\v{c}}n{\`y}}}, \bibinfo {author} {\bibfnamefont {L.}~\bibnamefont
  {Mike{\v{s}}}}, \bibinfo {author} {\bibfnamefont {A.}~\bibnamefont
  {Cecilia}}, \bibinfo {author} {\bibfnamefont {E.}~\bibnamefont {Hamann}},
  \bibinfo {author} {\bibfnamefont {T.}~\bibnamefont {Baumbach}}, \bibinfo
  {author} {\bibfnamefont {L.}~\bibnamefont {{\v{S}}v{\'e}da}}, \bibinfo
  {author} {\bibfnamefont {Z.}~\bibnamefont {Z{\'a}pra{\v{z}}n{\`y}}}, \bibinfo
  {author} {\bibfnamefont {D.}~\bibnamefont {Koryt{\'a}r}}, \bibinfo {author}
  {\bibfnamefont {E.}~\bibnamefont {Gimenez-Navarro}}, \emph {et~al.},\
  }\bibfield  {title} {\bibinfo {title} {Single-distance phase retrieval
  algorithm for bragg magnifier microscope},\ }\href
  {https://doi.org/https://doi.org/10.1364/OE.24.027753} {\bibfield  {journal}
  {\bibinfo  {journal} {Optics Express}\ }\textbf {\bibinfo {volume} {24}},\
  \bibinfo {pages} {27753} (\bibinfo {year} {2016})}\BibitemShut {NoStop}%
\bibitem [{\citenamefont {Spiecker}()}]{SpieckerPhD}%
  \BibitemOpen
  \bibfield  {author} {\bibinfo {author} {\bibfnamefont {R.}~\bibnamefont
  {Spiecker}},\ }\href {https://doi.org/10.5445/IR/1000162889} {}\bibinfo
  {note} {Dissertation, in preparation,
  \url{10.5445/IR/1000162889}}\BibitemShut {NoStop}%
\bibitem [{\citenamefont {Douissard}\ \emph {et~al.}(2012)\citenamefont
  {Douissard}, \citenamefont {Cecilia}, \citenamefont {Rochet}, \citenamefont
  {Chapel}, \citenamefont {Martin}, \citenamefont {van~de Kamp}, \citenamefont
  {Helfen}, \citenamefont {Baumbach}, \citenamefont {Luquot}, \citenamefont
  {Xiao} \emph {et~al.}}]{douissard2012versatile}%
  \BibitemOpen
  \bibfield  {author} {\bibinfo {author} {\bibfnamefont {P.-A.}\ \bibnamefont
  {Douissard}}, \bibinfo {author} {\bibfnamefont {A.}~\bibnamefont {Cecilia}},
  \bibinfo {author} {\bibfnamefont {X.}~\bibnamefont {Rochet}}, \bibinfo
  {author} {\bibfnamefont {X.}~\bibnamefont {Chapel}}, \bibinfo {author}
  {\bibfnamefont {T.}~\bibnamefont {Martin}}, \bibinfo {author} {\bibfnamefont
  {T.}~\bibnamefont {van~de Kamp}}, \bibinfo {author} {\bibfnamefont
  {L.}~\bibnamefont {Helfen}}, \bibinfo {author} {\bibfnamefont
  {T.}~\bibnamefont {Baumbach}}, \bibinfo {author} {\bibfnamefont
  {L.}~\bibnamefont {Luquot}}, \bibinfo {author} {\bibfnamefont
  {X.}~\bibnamefont {Xiao}}, \emph {et~al.},\ }\bibfield  {title} {\bibinfo
  {title} {A versatile indirect detector design for hard {X}-ray
  microimaging},\ }\href {https://doi.org/10.1088/1748-0221/7/09/P09016}
  {\bibfield  {journal} {\bibinfo  {journal} {Journal of Instrumentation}\
  }\textbf {\bibinfo {volume} {7}}\bibinfo  {number} { (09)},\ \bibinfo {pages}
  {P09016}}\BibitemShut {NoStop}%
\bibitem [{\citenamefont {Vogelgesang}\ \emph {et~al.}(2016)\citenamefont
  {Vogelgesang}, \citenamefont {Farago}, \citenamefont {Morgeneyer},
  \citenamefont {Helfen}, \citenamefont {dos Santos~Rolo}, \citenamefont
  {Myagotin},\ and\ \citenamefont {Baumbach}}]{vogelgesang2016real}%
  \BibitemOpen
\bibfield  {number} {  }\bibfield  {author} {\bibinfo {author} {\bibfnamefont
  {M.}~\bibnamefont {Vogelgesang}}, \bibinfo {author} {\bibfnamefont
  {T.}~\bibnamefont {Farago}}, \bibinfo {author} {\bibfnamefont {T.~F.}\
  \bibnamefont {Morgeneyer}}, \bibinfo {author} {\bibfnamefont
  {L.}~\bibnamefont {Helfen}}, \bibinfo {author} {\bibfnamefont
  {T.}~\bibnamefont {dos Santos~Rolo}}, \bibinfo {author} {\bibfnamefont
  {A.}~\bibnamefont {Myagotin}},\ and\ \bibinfo {author} {\bibfnamefont
  {T.}~\bibnamefont {Baumbach}},\ }\bibfield  {title} {\bibinfo {title}
  {{Real-time image-content-based beamline control for smart 4D {X}-ray
  imaging}},\ }\href
  {https://doi.org/https://doi.org/10.1107/S1600577516010195} {\bibfield
  {journal} {\bibinfo  {journal} {Journal of Synchrotron Radiation}\ }\textbf
  {\bibinfo {volume} {23}},\ \bibinfo {pages} {1254} (\bibinfo {year}
  {2016})}\BibitemShut {NoStop}%
\end{thebibliography}%

\clearpage

\appendix
\onecolumngrid

\section*{Supplementary Material}
\renewcommand\thefigure{S\arabic{figure}}    
\setcounter{figure}{0}  
\renewcommand\thetable{S\arabic{table}}

\section{Shift (in)variance}

\begin{figure*}[h]
\begin{center}
\includegraphics[scale=1.0]{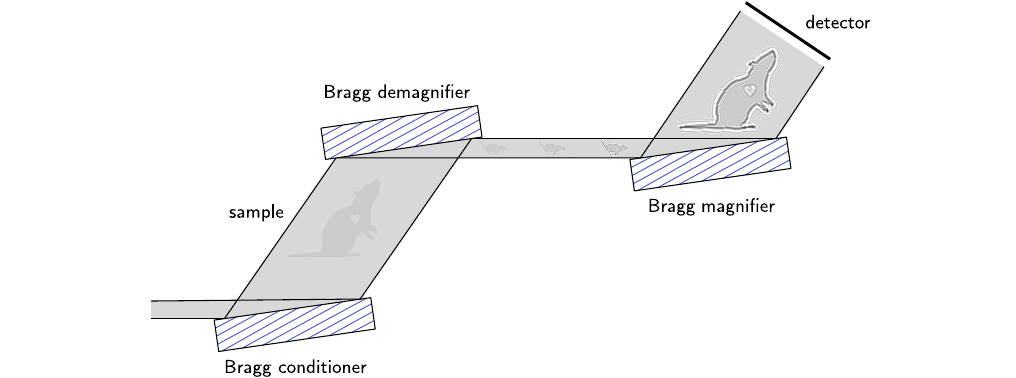}
\caption{Schematic of the combination Bragg conditioner, Bragg demagnifier and Bragg magnifier in 1D. The Bragg magnifier allows the usage of an efficient large-area detector. Further, when using the same type of crystals, the Bragg magnifier compensates the shift-variance of the demagnifier, see text.
} \label{fig_appendix}
\end{center}
\end{figure*}

In a shift-invariant imaging system, a translation of the object yields the same image, besides a displacement.
In contrast, Bragg crystal optics are in general shift-variant imaging systems~\cite{Modregger2008Mar,hrivvnak2018fast}. Due to asymmetric reflection a lateral translation of the wavefield leads to different propagation distances, which is a complication for phase retrieval~\cite{vagovivc2014x,hrivvnak2016single}. For coplanar diffraction of a plane wave on a crystal surface, the relation between the incident and outgoing spatial frequencies $k_\text{in}$ and $k_\text{out}$ is given by~\cite{SpieckerPhD}
\begin{align*}
		k_\text{out}(k_\text{in})=k_0\sin\left(\beta_\text{out} - \arccos\left[\cos\left(\beta_\text{in} - \arcsin \frac{k_\text{in}}{k_0}\right) + \cos\beta_\text{out} - \cos\beta_\text{in}\right]\right),
\end{align*}
where $\beta_\text{in}$ and $\beta_\text{out}$ are the angles between the crystal surface and the incident or outgoing optical axis, respectively. These angles can be computed from dynamical diffraction theory~\cite{authier2004dynamical}.
As mentioned in the main text, the shift-invariance of the demagnifier can be restored by adding a matched Bragg magnifier downstream (see \fref{fig_appendix}). 
This can be best understood in Fourier space. A plane wave with spatial frequency $\mathbf{k}_\text{in}$ incident on the demagnifier results in a plane wave with the same spatial frequency $\mathbf{k}_\text{out} = \mathbf{k}_\text{in}$ behind the magnifier. Only the phase and amplitude are altered by the crystals and the free space propagation. Consequently, the modulation of the propagated wavefield can be described by a transfer function, i.e. the propagated wavefield is given by a convolution of the incident wavefield with a coherent point-spread function. 

\section{Details on the experimental setup}
The experimental data were acquired at the beamline P23 at PETRA III, DESY, Hamburg. All four Si crystals used in the experiment have dimensions of \SI{21}{cm} x \SI{8}{cm} x \SI{2}{cm} and a nominal asymmetry angle of $\pm\SI{5.92}{\degree}$. The calculated imaging parameters of the Bragg demagnifier crystals are given in Table~\ref{tab_demag} for energies between \SI{29}{keV} to \SI{31}{keV}. Each crystal was mounted on a high-precision hexapod (Physik Instrumente (PI) GmbH \& Co. KG, 76228 Karlsruhe, Germany). For acquiring images with the Bragg demagnifier, we used an indirect detector system with a \SI{24}{\micro m} thick LSO:Tb scintillator, a diffraction-limited optical microscope (Optique Peter, 69210 Lentilly, France), a 5x objective lens with a numerical aperture of 0.14 (model 378-803-3, Mitutoyo Deutschland GmbH, 41469 Neuss, Germany), a \SI{180}{mm} tubus lens resulting in 4.5x total optical magnification \cite{douissard2012versatile}, and a pco.edge 5.5 CMOS camera (PCO AG, 93309 Kelheim, Germany) with an exposure time of \SI{67}{ms} and a physical pixel size of \SI{6.5}{\micro m} x \SI{6.5}{\micro m}, yielding an effective pixel size of \SI{1.44}{\micro m} in the image plane. Combined with the demagnification factor of $M=25.7$ at \SI{29}{keV}, the pixel size amounts to \SI{37}{\micro m} in the object plane.
For the conventional PB-PCI acquisitions, a Shad-o-Box 1K HS (Teledyne Dalsa, Waterloo, Canada) was placed \SI{1}{m} downstream of the sample. To prevent saturation of the camera, we had to attenuate the beam by a factor of 500. The presented images for conventional PB-PCI were acquired with an exposure time of \SI{67}{ms} averaged over 20 acquisitions. For data acquisition and motor control, we used the control system \textit{Concert}~\cite{vogelgesang2016real}. The tiger salamander (\textit{Ambystoma tigrinum}) utilized in the experiment is a museum specimen that was loaned for our experiment. The mouse liver was obtained from The Jackson Laboratory, US. Both specimens were fixed in PFA and subsequently stored in \SI{70}{\percent} ethanol.

\begin{table}[h]
	\begin{center}
		\caption{Calculated parameters for the Si (220) Bragg demagnifier with an asymmetry angle of \SI{-5.92}{\degree}.}
		\label{tab_demag}
		\begin{tabular}{c|c|c|c|c|c}
			energy (keV) & resolution limit ($\si{\micro m}$) & demagnification $M$ & gain factor $M^2$ \\
			
			\hline
			29.0 &  68 &  25.7 & 660\\
			30.0 &  91 &  45.4 & 2061\\
			30.5 &  116 &  71.4 & 5098\\
			31.0 &  190 &  144.7 & 20938			
		\end{tabular}
	\end{center}
\end{table}

\end{document}